\documentclass[a4paper,11pt]{article}
\usepackage{jcappub}
\usepackage{aas_macros}

\newcommand{\nv}{\hat{\bf n}}
\newcommand{\bU}[1]{\langle bU_{\rm #1}\rangle}

\title{Detailed theoretical modelling of the kinetic Sunyaev-Zel'dovich stacking power spectrum}

\author{Amy Wayland$^1$,}
\author{David Alonso$^1$,}
\author{and Adrien La Posta$^1$}
\affiliation{$^1$Department of Physics, University of Oxford, Denys Wilkinson Building, Keble Road, Oxford OX1 3RH, United Kingdom}

\emailAdd{amy.wayland@physics.ox.ac.uk}

\abstract{We examine, from first principles, the angular power spectrum between the kinematic Sunyaev-Zel'dovich effect (kSZ) and the reconstructed galaxy momentum -- the basis of existing and future ``kSZ stacking'' analyses. We present a comprehensive evaluation of all terms contributing to this cross-correlation, including both the transverse and longitudinal modes of the density-weighted velocity field, as well as all irreducible correlators that contribute to the momentum power spectrum. This includes the dominant component, involving the convolution of the electron-galaxy and velocity-velocity power spectra, an additional disconnected cross-term, and a connected non-Gaussian trispectrum term. Using this framework, we examine the impact of other commonly neglected contributions, such as the two-halo component of the dominant term, and the impact of satellite galaxies. Finally, we assess the sensitivity of upcoming CMB experiments to these effects and determine that they will be sensitive to the cross-term, the connected non-Gaussian trispectrum term, the two-halo contribution and impact of satellite galaxies, at a significance level of $\sim4-6\sigma$. On the other hand, the contribution from longitudinal modes is negligible in all cases. These results identify the astrophysical observables that must be accurately modelled to obtain unbiased constraints on cosmology and astrophysics from near-future kSZ measurements.}

\begin{document}
\maketitle
\flushbottom

\section{Introduction}
  The scattering of Cosmic Microwave Background (CMB) photons by hot electrons in galaxy clusters induces anisotropies in the observed CMB temperature, a phenomenon known as the Sunyaev-Zel'dovich (SZ) effect \cite{1969Ap&26SS...4..301Z,Sunyaev:1970,Sunyaev:1972,1811.02310,astro-ph/9808050,Bolliet:2022}. This effect can be decomposed into two components: the thermal Sunyaev-Zel'dovich (tSZ) effect, arising from the thermal motion of free electrons in the intra-cluster and inter-galactic media (ICM and IGM), and the kinetic Sunyaev-Zel'dovich (kSZ) effect, caused by the Doppler shift of CMB photons due to their bulk velocity relative to the CMB rest frame. Although both effects are secondary anisotropies of the CMB, the distinct frequency dependence of the tSZ allows it to be separated from the CMB, kSZ, and other sky components in multi-frequency analyses. In contrast, the kSZ effect preserves the blackbody frequency spectrum of the CMB and therefore cannot be independently separated from the primary CMB anisotropies, which then act as a source of noise that dominates on large angular scales. 
  
  Observationally, the late-time kSZ signal can instead be extracted by cross-correlating CMB temperature maps with the positions of galaxies and other tracers of the large-scale structure (LSS), weighted by an estimate of their radial velocities. This is the general principle of the so-called ``kSZ tomography'' approach \cite{1810.13423}, which has seen different implementations, in the form of the pairwise kSZ estimator \cite{astro-ph/9812456,1203.4219,2101.08374,2307.11894}, large-scale velocity reconstruction \cite{1810.13424,2410.06229,2411.08240,2506.21657}, cross-correlation with angular redshift fluctuations \cite{1911.10690}, and kSZ stacking \cite{1510.06442,Schaan:2021,Hadzhiyska:2024}. Other procedures have also been proposed, including more general bispectrum estimators \cite{1810.13423,2411.11974} and projected fields \cite{1603.01608,2102.01068}. An excellent review of these methods can be found in \cite{Bolliet:2022}.

  kSZ measurements have become an invaluable tool for both cosmology and astrophysics. Their sensitivity to the cosmic velocity field could allow them to constrain the growth rate of structure \cite{1408.6248,1604.01382,1712.05714}, assuming that the optical depth of the sample under study can be characterised independently. The sensitivity of the velocity field to the large-scale matter distribution could also make kSZ cross-correlations a highly-sensitive probe of primordial non-Gaussianity \cite{1810.13424,2410.06229,2411.08240,2506.21657}. In turn, their sensitivity to the distribution of ionised gas has transformed kSZ measurements into a potentially crucial tool to improve our understanding of feedback processes, and their repercussion on the distribution of gas and the corresponding suppression in the small-scale clustering of matter (of key importance for cosmological weak lensing measurements) \cite{Hadzhiyska:2024,2404.06098,2410.19905,2505.20413,2509.10455}.  With new CMB experiments, such as the Simons Observatory \cite{SO:2018,2503.00636}, with enhanced sensitivity, measurements of the kSZ effect will swiftly grow in precision, increasing their potential to address these questions.

  Here, we will focus on kSZ stacking measurements, aimed at recovering the small-scale distribution of gas. The astrophysical models currently used to describe this signal are relatively simple, owing to the modest sensitivity with which the kSZ has been detected so far. In particular, stacked kSZ measurements are often interpreted under the assumption that the signal arises solely from the line-of-sight component of the electron momentum field within individual haloes, often neglecting several important contributions. These include the two-halo term arising from correlated large-scale structure, the impact of miscentering and satellite galaxies in observational galaxy and cluster catalogues, and higher-order correlations between the velocity and density fields. The increased sensitivity of future experiments may make them sensitive to some of these effects, which must therefore be incorporated in the theoretical models used to interpret them. To address this, this work aims to present a comprehensive theoretical description of the kSZ effect, by analytically deriving the full angular power spectrum from first principles, incorporating contributions from both transverse and longitudinal modes of the density-weighted velocity field, and accounting for the correlation between density and velocity fields at all orders. Our calculations will build on previous work presented in the literature, including \cite{Ma:2001,1109.0553,Park:2013,Bolliet:2022,2306.03127}, many of which focused on the kSZ auto-spectrum or the projected-fields estimator, applying it to the case of kSZ stacking.

  The structure of this paper is as follows. In Section \ref{sec:theory}, we derive analytical expressions for the contributions of both transverse and longitudinal density-weighed velocity modes to the kSZ-galaxy angular power spectrum, along with the corresponding 3D power spectra. We provide a complete evaluation of all terms contributing to the angular power spectrum, which will be essential for precise kSZ stacking analyses. In Section \ref{sec:results}, we numerically implement our analytical framework to quantify the relative magnitudes of the various terms contributing to the kSZ-galaxy power spectrum. Additionally, we investigate the impact of including the two-halo term and satellite galaxies in the Halo Occupation Distribution (HOD) model. We further examine the sensitivity of upcoming CMB surveys to these different contributions, as well as the impact of the choice of baryon model and velocity reconstruction. Finally, we conclude in Section \ref{sec:conclusions} with a summary of our key findings and a discussion of their implications for future CMB experiments.
 
\section{Theory} \label{sec:theory}
  Mathematically, the signal from the kSZ effect at low redshifts can be detected by cross-correlating the projected galaxy momentum field, $\pi^{\rm g}$, with maps of the CMB temperature fluctuations. The galaxy momentum field is expressed as
  \begin{equation} \label{eq:pi_g}
    \pi^{\rm g}(\nv) = \int \mathrm{d}\chi \, H(z) p(z) (1+\delta_{\rm g}(\chi\nv)) \nv \cdot \mathbf{v},
  \end{equation}
  where $\nv$ is a line-of-sight unit vector, ${\bf v}$ is the peculiar velocity field, $\delta_{\rm g}({\bf x})$ is the overdensity of galaxies, $p(z)$ is the redshift distribution of the galaxy sample under study, and $H(z)$ is the expansion rate at redshift $z$\footnote{Note that we use natural units, with $c=1$, throughout.}. It is important to note that, in practice, the velocity field in Equation \eqref{eq:pi_g} corresponds to the reconstructed velocity rather than the true velocity. This distinction will be revisited in Section \ref{sssec:results.sensitivity.velrec}, where we explore the assumption of using the true velocity as opposed to the reconstructed velocity.

  The temperature shift of the CMB due to the kSZ effect, $\Delta T$, is in turn given by
  \begin{equation} \label{eq:Delta_T}
    \left.\frac{\Delta T}{T}\right|_{\rm kSZ}(\nv) = \sigma_T \int \frac{\mathrm{d}\chi}{1+z} \, e^{-\tau} \bar{n}_{\rm e}(z) (1+\delta_{\rm e}(\chi\nv)) \nv \cdot \mathbf{v}.
  \end{equation}
  Here, $\sigma_T$ is the Thomson scattering cross section, $\bar{n}_{\rm e}$ is the mean physical number density of free electrons, and $\delta_{\rm e}$ is the electron overdensity. 
  
  Both $\Delta T/T$ and $\pi^{\rm g}$ can be written as a generic projected momentum field of the form
  \begin{equation}
    \pi^{\rm a}(\nv)\equiv \int d\chi\,W_{\rm a}(\chi)\,{\bf q}_{\rm a}(\chi\nv,z)\cdot\nv,
  \end{equation}
  where $W_{\rm a}(\chi)$ is a radial kernel and ${\bf q}_{\rm a}({\bf x},z)\equiv(1+\delta_{\rm a}({\bf x},z))\,{\bf v}$ is a 3D momentum field. To denote the galaxy momentum and the kSZ field, we will use the subscripts ${\rm a}={\rm g}$ and ${\rm a}={\rm e}$, respectively. Then, the radial kernels are
  \begin{equation}
    W_{\rm g}(\chi)=H(z)\,p(z),\hspace{12pt}
    W_{\rm e}(\chi)=\sigma_T\bar{n}_{{\rm e},0}(1+z)^2,
  \end{equation}
  where $\bar{n}_{{\rm e},0}$ is the mean electron number density today. The angular power spectrum associated with the kSZ effect is then obtained from the two-point correlator of the spherical harmonic coefficients of the fields $\pi^{\rm g}$ and $\pi^{\rm e}\equiv\Delta T/T|_{\rm kSZ}$. The resulting angular power spectrum is then defined as
  \begin{equation}
    \langle \pi^{\rm a}_{\ell m}\pi^{{\rm b}*}_{\ell'm'}\rangle \equiv \delta^K_{\ell\ell'}\delta^K_{mm'}C_\ell^{\rm ab},
  \end{equation}
  where $\delta^K_{ij}$ is the Kronecker delta function.
  
  Note that the popular ``kSZ stacking'' estimator \cite{1510.06442,Schaan:2021} used to study the small-scale gas distribution can be expressed as a particular convolution of $C_\ell^{\rm ge}$ with the window function of the compensated aperture photometry filters used in the estimator \cite{LKDA_inprep}. For simplicity, and without loss of generality, we will simply express our calculation in terms of the angular power spectrum, ensuring that we recover all information available in the data.

  As described in \cite{Ma:2001,Park:2013,Bolliet:2022}, the dominant contribution to $C_\ell^{\rm ge}$ comes from the transverse components of the 3D momentum fields (i.e. Fourier modes of ${\bf q}_{\rm a}$ that are perpendicular to the wave vector ${\bf k}$). For completeness, this work will calculate the contribution from both transverse and longitudinal modes, but we will start with the former. We will do so following the techniques outlined in \cite{Park:2013}.

  \subsection{Power spectra for transverse modes} \label{ssec:theory.perp}
    The transverse momentum mode is defined in Fourier space as
    \begin{equation}
      {\bf q}^\perp_{\rm a}({\bf k})\equiv{\bf q}_{\rm a}({\bf k})-[\hat{\bf k}\cdot{\bf q}_{\rm a}({\bf k})]\hat{\bf k},
    \end{equation}
    where $\hat{\bf k}\equiv{\bf k}/k$ is a unit vector in the direction of the wavenumber ${\bf k}$. The contribution from this transverse mode to the projected momentum field is
   \begin{equation} \label{eq:pi_perp_FT}
       \pi^{\rm a}_{\perp}(\nv) = \int \mathrm{d}\chi \, W_{\rm a}(\chi) \int \frac{\mathrm{d}^3 k}{(2\pi)^3} \, \cos(\phi_{\hat{q}_{\rm a}}-\phi_{\hat{n}})(1-x^2)^{1/2} \, q_{\rm a}^{\perp}(\mathbf{k},z) e^{-ik\chi\,x}
   \end{equation}
   where $x = \hat{\mathbf{k}} \cdot \nv$, and $\phi_{\hat{q}_{\rm a}}$ and $\phi_{\hat{n}}$ denote the azimuthal angles of the directions $\hat{\mathbf{q}}_{\rm a}$ and $\nv$, respectively.
  
   \subsubsection{Angular power spectrum} \label{sssec:theory.perp.C_ell}
     We begin by computing the spherical harmonic coefficients of $\pi^{\rm a}_\perp$ as defined in Equation \eqref{eq:pi_perp_FT}:
     \begin{align}
       \pi^{\rm a}_{\perp,{\ell m}} &= \int \mathrm{d}^2 \nv \, Y_{\ell}{}^{m^*}(\nv) \pi^{\rm a}_{\perp}(\nv).
     \end{align}
     Following the steps outlined in \cite{Park:2013}, we obtain the spherical harmonic coefficients
     \begin{align}\label{eq:pilm_perp}
        \pi^{\rm a}_{\perp,{\ell m}} &= A_{\ell} \int \frac{\mathrm{d}^3 k}{(2\pi)^3} \, \sum_{m'=\pm1} (-i)^{\ell+1} \mathsf{D}_{m,m'}^{\ell}(S(\hat{\mathbf{k}})) \int \mathrm{d}\chi \, W_{\rm a}(\chi)\,q_{\rm a}^\perp(\mathbf{k},\chi) \frac{j_{\ell}(k\chi)}{k\chi},
    \end{align}
    where the $\ell$-dependent prefactor is given by $A_\ell \equiv \sqrt{\pi \ell(\ell+1)(2\ell+1)}$, $j_{\ell}$ is the spherical Bessel function of order $\ell$, $S(\hat{\mathbf{k}})$ is the rotation operator that maps the $z$-axis to the direction $\hat{\mathbf{k}}$, and $\mathsf{D}_{m,m'}^{\ell} = \left\langle \ell,m' |S| \ell,m \right\rangle$ denotes the Wigner-$\mathsf{D}$ matrix elements, representing the action of this rotation on the spherical harmonics.

    To compute the angular cross-power spectrum, we take the two-point correlator of the spherical harmonic coefficients. Using the orthogonality of the Wigner-$\mathsf{D}$ matrices, we then obtain the following expression:
    \begin{equation}
        C^{{\rm ab},\perp}_\ell = \frac{\ell(\ell+1)}{\pi} \int \mathrm{d}\chi\,W_{\rm a}(\chi) \int \mathrm{d}\chi'\,W_{\rm b}(\chi') \int \mathrm{d}k \, k^2 \frac{j_{\ell}(k\chi)}{k\chi} \frac{j_{\ell}(k\chi')}{k\chi'} P^{\rm ab}_\perp(k,z),
    \end{equation}
    where the 3D power spectrum for the transverse momentum mode is defined as
    \begin{equation}
      \left\langle \left[{\bf q}_{\rm a}^{\perp}(\mathbf{k})\right]^\dagger {\bf q}_{\rm b}^{\perp}(\mathbf{k}') \right\rangle = (2\pi)^3\,\delta^{\cal D}(\mathbf{k}-\mathbf{k}')\,P^{\rm ab}_{\perp}(k).
    \end{equation} 
    This expression can be further simplified by applying the Limber approximation, which consists of replacing the Bessel functions by
    \begin{equation}\label{eq:limber}
      j_\ell(x)\simeq\sqrt{\frac{\pi}{2\ell+1}}\delta^D\left(\ell+\frac{1}{2}-x\right).
    \end{equation}
    The result is\footnote{This neglects an $\ell$-dependent prefactor $\ell(\ell+1)/(\ell+1/2)^2$, which is close to 1 on all relevant scales.}:
    \begin{equation} \label{eq:C_ell_perp}
      C^{{\rm ab},\perp}_\ell = \frac{1}{2} \int \frac{\mathrm{d}\chi}{\chi^2}\,W_{\rm a}(\chi)\,W_{\rm b}(\chi)\,P^{\rm ab}_\perp\left(k=\frac{\ell+1/2}{\chi}, \, z\right).
    \end{equation}
   
   \subsubsection{3D power spectrum} \label{sssec:theory.perp.3D}
     In order to compute the angular power spectrum from Equation \eqref{eq:C_ell_perp}, we begin by calculating the 3D power spectrum, $P_{\perp}^{\rm ab}(k)$, which describes the power spectrum of the curl of the momentum field. This requires first determining the Fourier transform of the momentum field, given by
     \begin{equation} \label{eq:q_FT}
       \mathbf{q}_{\rm a}(\mathbf{k}) = \mathbf{v}(\mathbf{k}) + \int \frac{\mathrm{d}^3 k'}{(2\pi)^3} \, \delta_{\rm a}(\mathbf{k}-\mathbf{k}')\, \mathbf{v}(\mathbf{k}').
     \end{equation}
     In what follows, we will assume that the velocity field can be expressed in terms of the matter density fluctuations, $\delta_{\rm m}$, using linear theory:
     \begin{equation} \label{eq:vel_linear_theory}
       \mathbf{v}(\mathbf{k}) = aHf \frac{i\mathbf{k}}{k^2} \delta_{\rm m}(\mathbf{k}),
     \end{equation}
     where $H$ is the expansion rate, and $f = \mathrm{d} \ln D/\mathrm{d}\ln a$ is the growth rate, given by the logarithmic derivative of the linear growth factor, $D$. Equation \eqref{eq:vel_linear_theory} is the linearised continuity equation, which neglects higher-order non-linear contributions present in the full Newtonian perturbation equations. Additionally, we omit the rotational component of the velocity field, which is not captured in the scalar formulation used here. The impact of these contributions will be quantified in future work.
     
     Substituting Equation \eqref{eq:vel_linear_theory} into Equation \eqref{eq:q_FT}, we obtain
     \begin{equation}
       \mathbf{q}_{\rm a}(\mathbf{k}) = i\,aHf \left[\frac{\mathbf{k}}{k^2}\delta_{\rm m}(\mathbf{k}) + \int \frac{\mathrm{d}^3 k'}{(2\pi)^3} \, \frac{\mathbf{k}'}{(k')^2} \delta_{\rm a}(\mathbf{k}-\mathbf{k}') \delta_{\rm m}(\mathbf{k'}) \right].
     \end{equation}
     To isolate the transverse component, we project the result onto the plane orthogonal to $\mathbf{k}$, which yields
     \begin{equation} \label{eq:q_FT_perp}
       \mathbf{q}_{\rm a}^{\perp}(\mathbf{k}) = i\,aHf \int \frac{\mathrm{d}^3 k'}{(2\pi)^3} \left(\frac{\mathbf{k}'}{(k')^2} - \frac{\mathbf{k}(\mathbf{k}\cdot\mathbf{k}')}{k^2(k')^2}\right) \delta_{\rm a}(\mathbf{k}-\mathbf{k}') \delta_{\rm m}(\mathbf{k}').
     \end{equation}
     The 3D power spectrum of the perpendicular component, $P_{q_\perp}^{\rm ge}(k)$, is then given by
     \begin{align}\nonumber
       \left\langle \left[{\bf q}_{\rm g}^{\perp}(\mathbf{k})\right]^\dagger {\bf q}_{\rm e}^{\perp}(\mathbf{k}) \right\rangle = \frac{(aHf)^2}{(2\pi)^3} \int \frac{\mathrm{d}^3k'\,\mathrm{d}^3k''}{(2\pi)^6} \frac{{\bf k}'_\perp\cdot{\bf k}''_\perp}{(k')^2(k'')^2} \langle \delta_{\rm g}(\mathbf{k}-\mathbf{k}') \delta_{\rm m}(\mathbf{k}')\delta_{\rm e}(-\mathbf{k}+\mathbf{k}'') \delta_{\rm m}(-\mathbf{k}'')\rangle,
    \end{align}
    where we have defined ${\bf k}'_\perp \equiv \mathbf{k}' - \mathbf{k}\,(\mathbf{k}\cdot\mathbf{k}')/k^2$. By decomposing the four-point correlator into its connected and disconnected components, we identify three contributions to the total 3D power spectrum:
    \begin{align}
        P^{\rm ge}_{\perp,1}(k) &= (aHf)^2 \int \frac{\mathrm{d}^3 k'}{(2\pi)^3}\, \frac{|\mathbf{k}'_\perp|^2}{(k')^4} P_{\rm ge}(|\mathbf{k}-\mathbf{k}'|) P_{\rm mm}(k'), \\[0.5em]
        P^{\rm ge}_{\perp,2}(k) &= - (aHf)^2 \int \frac{\mathrm{d}^3k'}{(2\pi)^3} \frac{(1-(\mu')^2)}{|\mathbf{k}-\mathbf{k}'|^2} P_{\rm gm}(|\mathbf{k}-\mathbf{k}'|) P_{\rm em}(k'), \\[0.5em]
        P^{\rm ge}_{\perp, \rm c}(k) &= (aHf)^2 \int \frac{\mathrm{d}^3 k'\,\mathrm{d}^3 k''}{(2\pi)^6} \, \frac{\mu_{12} - \mu' \mu''}{k' k''} T_{\rm gemm}(\mathbf{k} - \mathbf{k}', -(\mathbf{k} - \mathbf{k}''), \mathbf{k}', -\mathbf{k}''),
    \end{align}
    such that the total perpendicular power spectrum is given by $P^{\rm ge}_\perp = P^{\rm ge}_{\perp,1} + P^{\rm ge}_{\perp,2} + P^{\rm ge}_{\perp, \rm c}$. Here, we have denoted the $\langle \rm ge \rangle \langle mm \rangle$ and $\langle \rm gm \rangle \langle em \rangle$ terms as $P_{\perp,1}^{\rm ge}$ and $P_{\perp,2}^{\rm ge}$, respectively, for ease of notation. Additionally, the directional cosines are defined as $\mu' = \hat{\mathbf{k}} \cdot \hat{\mathbf{k}}'$, $\mu'' = \hat{\mathbf{k}} \cdot \hat{\mathbf{k}}''$, and $\mu_{12} = \hat{\mathbf{k}}' \cdot \hat{\mathbf{k}}''$, and we define the power spectra and connected trispectrum via
    \begin{align}
      &\langle \delta_{\rm a}(\mathbf{k}_1) \delta_{\rm b}(\mathbf{k}_2) \rangle \equiv (2\pi)^3 \delta^{\mathcal{D}}(\mathbf{k}_1 + \mathbf{k}_2) P_{\delta_{\rm A} \delta_{\rm B}}(\mathbf{k}_1),\\
      &\langle \delta_{\rm a}(\mathbf{k}_1) \delta_{\rm b}(\mathbf{k}_2) \delta_{\rm c}(\mathbf{k}_3) \delta_{\rm d}(\mathbf{k}_4)\rangle_{\mathrm{c}} \equiv (2\pi)^3 \delta^{\mathcal{D}}(\mathbf{k}_1 + \mathbf{k}_2 + \mathbf{k}_3 + \mathbf{k}_4) T_{\rm abcd}(\mathbf{k}_1, \mathbf{k}_2, \mathbf{k}_3, \mathbf{k}_4).
    \end{align}
    Rewriting the expressions above in terms of spherical coordinates for the internal wave vectors, defined with the external mode ${\bf k}$ aligned along the $z$ axis, they can be simplified to
    \begin{align}
      P^{\rm ge}_{\perp,1}(k) &= \frac{(aHf)^2}{(2\pi)^2} \int \mathrm{d}k' \, P_{\rm mm}(k') \int_{-1}^{1} \mathrm{d}\mu' \, (1-(\mu')^2) P_{\rm ge} (|\mathbf{k}-\mathbf{k}'|), \label{eq:P_perp_1} \\[0.5em]
      P^{\rm ge}_{\perp,2}(k) &= -\frac{(aHf)^2}{(2\pi)^2} \int \mathrm{d}k' \, (k')^2 P_{\rm em}(k') \int_{-1}^{1} \mathrm{d}\mu' \, (1-(\mu')^2) \frac{P_{\rm gm}(|\mathbf{k}-\mathbf{k}'|)}{|\mathbf{k}-\mathbf{k}'|^2}, \label{eq:P_perp_2} \\[0.5em]
      P^{\rm ge}_{\perp, \rm c}(k) &= \frac{(aHf)^2}{(2\pi)^5} \int \mathrm{d}k' \, k' \int \mathrm{d}k'' \, k'' \int_{-1}^1 \mathrm{d}\mu' \, \sqrt{1-(\mu')^2} \int_{-1}^1 \mathrm{d}\mu'' \, \sqrt{1-(\mu'')^2} \int_0^{2\pi} \mathrm{d}\phi \, \cos\phi \nonumber \\[0.2em] 
      &\hspace{40pt}\times T_{\rm gemm}(\mathbf{k} - \mathbf{k}', -\mathbf{k} + \mathbf{k}'', \mathbf{k}', -\mathbf{k}''), \label{eq:P_perp_c}
    \end{align}
    where $\phi = \phi' - \phi''$.

    \subsubsection{The OV limit}\label{sssec:theory.perp.OV}
      As a consistency test, we can recover the expression for the Ostriker-Vishniac (OV) power spectrum \citep{Ostriker:1986, Vishniac:1987} in the limit where all overdensities ($\delta_{\rm g},\,\delta_{\rm e},\,\delta_{\rm m}$) are replaced by the matter overdensity in the linear regime, $\delta_{\rm lin}$. By summing the contributions from Equations \eqref{eq:P_perp_1} and \eqref{eq:P_perp_2} in this limit, we obtain
      \begin{equation}
        P_{\perp}^{\rm OV}(k) = (aHf)^2 \int \frac{\mathrm{d}^3 k'}{(2\pi)^3} \left(\frac{k(k - 2 k' \mu')(1-(\mu')^2)}{k^2 + (k')^2 - 2 k k' \mu'}\right) P_{\rm lin}(k') P_{\rm lin}(|\mathbf{k}-\mathbf{k}'|),
      \end{equation}
      where $P_{\rm lin}(k)$ is the linear matter power spectrum. This is consistent with the results presented in e.g. \cite{Ma:2001,Bolliet:2022}.

    \subsubsection{Small-scale limit and stacking}\label{sssec:theory.perp.hik}
      On small scales, the amplitude of the kSZ signal is primarily dominated by the $P_{\perp,1}(k)$ term \citep{astro-ph/9907103,Ma:2001,Park:2013,Bolliet:2022}, as we will also show later. In this regime (i.e. for large $k$), we may also approximate $P_{\rm ge}(|{\bf k}-{\bf k}'|)\simeq P_{\rm ge}(k)$ in Equation \eqref{eq:P_perp_1}, since the velocity power spectrum, $P_{\rm vv}(k)\propto P_{\rm mm}(k)/k^2$, is heavily weighted towards large scales (i.e. $k'\ll k$ always). Hence, the transverse power spectrum can be written approximately as
      \begin{equation}\label{eq:ptrans_hik}
        P_{\perp}^{\rm ge}(k)\simeq 2\sigma_{v_r}^2\,P_{\rm ge}(k),\hspace{6pt}{\rm with}\hspace{6pt} \sigma_{v_r}^2\equiv\frac{1}{3}\int \frac{\mathrm{d}k'}{2\pi^2}\,k'^2\,P_{\rm vv}(k'),
      \end{equation}
      where $\sigma_{v_r}^2$ is the radial velocity dispersion. 
      
      Furthermore, keeping only the one-halo contribution to the galaxy-electron power spectrum, and assuming that all galaxies in the sample are centrals, we obtain
      \begin{equation}
        P_\perp^{\rm ge}(k)\sim\frac{2\sigma_{v_r}^2}{\bar{n}_{\rm e}}\left\langle n_{\rm e}(k)\right\rangle,
      \end{equation}
      where $\langle n_{\rm e}(k)\rangle$ is the Fourier-space electron density profile averaged over the halo masses probed by the sample:
      \begin{equation}
        \left\langle n_{\rm e}(k)\right\rangle\equiv\frac{1}{\bar{n}_{\rm g}}\int \mathrm{d}M\,n(M)\,N_{\rm g}(M)\,n_{\rm e}(k|M),\hspace{12pt} \bar{n}_{\rm g}\equiv\int \mathrm{d}M\,n(M)\,N_{\rm g}(M),
      \end{equation}
      where $n_{\rm e}(k|M)$ is the Fourier-space electron profile for haloes of mass $M$, and $n(M)$ is the halo mass function. In this limit, we can therefore interpret the kSZ cross-correlation as providing a direct measurement of the gas density profile in the host haloes of the galaxy sample under study. This leads to the most common interpretation of the stacked kSZ signal \cite{1510.06442,Schaan:2021,2404.06098,2505.20413}. However, deviations from this interpretation arise due to contributions from a significant fraction of satellite galaxies in the sample, the two-halo term, and the subdominant terms $P_{\perp,2}^{\rm ge}(k)$ and $P_{\perp,{\rm c}}^{\rm ge}(k)$. These contributions may be relevant, particularly given the sensitivity of upcoming experiments. Quantifying their impact -- in addition to the contribution from longitudinal modes, which will be described in the next section -- is the core aim of this paper.

  \subsection{Power spectra for the longitudinal mode}\label{ssec:theory.par}
    We now consider the longitudinal mode of the Fourier transform of the momentum field. As before, we start by computing the longitudinal component of the projected galaxy momentum field and of the kSZ temperature shift:
    \begin{equation} \label{eq:pi_par_FT}
       \pi^{\rm a}_\parallel(\nv) = \int \mathrm{d}\chi \, W_{\rm a}(\chi) \int \frac{\mathrm{d}^3 k}{(2\pi)^3} \, x\,q_{\rm a}^{\parallel}(\mathbf{k},z) e^{-ik\chi x}.
   \end{equation}
   As noted in \cite{Park:2013}, cancellations caused by the oscillatory factor $\exp(-ik\chi x)$ suppress the integral over $\chi$ unless the product $kx$ is of the order of the horizon $H$ (which is the timescale over which ${\bf q}_{\rm a}(k,z)$ varies). For large $k$, where the kSZ signal can be detected, this implies that $x$ must be small or, in other words, that the most relevant Fourier modes are the transverse ones. We will nevertheless carry out the calculation of the longitudinal contribution to the kSZ stacking power spectra for completeness. This is not an entirely futile effort since, in the case of kSZ stacking against galaxies in a sufficiently narrow redshift bin (i.e. when $W_{\rm a}(\chi)$ varies over radial distances much smaller than the horizon), longitudinal contributions may become more relevant.

   \subsubsection{Angular power spectrum} \label{sssec:theory.par.C_ell}
     Following similar steps to Section \ref{sssec:theory.perp.C_ell}, we begin by computing the spherical harmonic decomposition of Equation \eqref{eq:pi_par_FT} via
     \begin{align}
       \pi^{\rm a}_{\parallel, \ell m} = \int \mathrm{d}^2 \nv \, Y_{\ell}{}^{m^*}(\nv) \pi^{\rm a}_\parallel(\nv).
     \end{align}
     In this case, we obtain the spherical harmonic coefficients
     \begin{align}
       \pi^{\rm a}_{\parallel,\ell m} = B_{\ell} \int \frac{\mathrm{d}^3 k}{(2\pi)^3} (-i)^{\ell+1} \mathsf{D}_{m,0}^{\ell}(S(\hat{\mathbf{k}})) \int \mathrm{d}\chi \, W_{\rm a}(\chi) q_{\rm a}^{\parallel}(\mathbf{k},\chi)\,j_{\ell}'(k\chi),
     \end{align}
     where the $\ell$-dependent prefactor is given by $B_{\ell} \equiv \sqrt{4\pi(2\ell+1)}$ and $j_{\ell}'$ is the first derivative of the spherical Bessel function of order $\ell$. Note that the main change with respect to the transverse modes (Equation \eqref{eq:pilm_perp}) is the dependence only on $m'=0$, instead of $m'\in\{-1,+1\}$, and the presence of $j_\ell'(x)$ instead of $j_\ell(x)/x$. These differences originate from the different angular prefactors entering Equations \eqref{eq:pi_perp_FT} and \eqref{eq:pi_par_FT}, and their harmonic expansion. Using this result, the longitudinal contribution to the kSZ power spectrum is
     \begin{equation}\label{eq:C_ell_par}
       C_\ell^{{\rm ab},\parallel} = \frac{2}{\pi} \int \mathrm{d}\chi\,W_{\rm a}(\chi) \int \mathrm{d}\chi'\,W_{\rm b}(\chi') \int \mathrm{d}k \, k^2 j_{\ell}'(k\chi) j_{\ell}'(k\chi') P^{\rm ab}_{\parallel}(k,z),
     \end{equation}
     where the 3D power spectrum of the longitudinal momentum mode is defined by
     \begin{equation}
       \langle q_{\rm a}^{\parallel}(\mathbf{k}) (q_{\rm b}^{\parallel}(\mathbf{k}'))^* \rangle \equiv (2\pi)^3 \delta^{\mathcal{D}}(\mathbf{k}-\mathbf{k}') P_{\parallel}^{\rm ab}(k).    
     \end{equation}
     
     To obtain an expression for $C_\ell^{{\rm ab},\parallel}$ in the Limber approximation, we may expand the Bessel function derivative as $j'_\ell(x)=(\ell/x)j_\ell(x)-j_{\ell+1}(x)$ and then apply the approximation in Equation \eqref{eq:limber}. The resulting expression is not visually simpler than the non-Limber result, and may be significantly less accurate for narrow radial kernels compared to the result obtained for transverse modes.

     We note that the derivation of the CMB angular power spectrum for the longitudinal mode, corresponding to the case $\mathrm{a} = \mathrm{b} = \mathrm{e}$, is presented in the appendix of \cite{Park:2016}. In the present work, however, we focus on the cross-correlation case.

   \subsubsection{3D power spectrum} \label{sssec:theory.par.3D}
     We now compute the 3D power spectrum of the longitudinal momentum mode, $P^{\rm ab}_{\parallel}(k)$, entering Equation \eqref{eq:C_ell_par}. In this case, we begin with the longitudinal component of the Fourier transform of the momentum field, which is given by
     \begin{equation}
       q_{\rm a}^{\, \parallel}(\mathbf{k}) = aHf \frac{\mathbf{k}}{k^2} \left[\delta_{\rm m}(\mathbf{k}) + \int \frac{\mathrm{d}^3 k'}{(2\pi)^3} \, \frac{\mathbf{k} \cdot \mathbf{k}'}{(k')^2} \delta_{\rm a}(\mathbf{k}-\mathbf{k}') \delta_{\rm m}(\mathbf{k}') \right].
    \end{equation}
    The corresponding power spectrum, $P_{q_{\parallel}}^{\rm ge}(k)$, is then
    \begin{equation}
      \langle q^\parallel_{\rm g}({\bf k})(q^\parallel_{\rm e}({\bf k}))^*\rangle = \frac{(aHf)^2}{k^2 (2\pi)^3} \langle \mathcal{D}_{\rm g}({\bf k}) \mathcal{D}_{\rm e}^*({\bf k}) \rangle,
    \end{equation}
    where 
    \begin{align}
       \mathcal{D}_{\rm a}({\bf k}) &\equiv \delta_{\rm m}(\mathbf{k}) + \int \frac{\mathrm{d}^3 k'}{(2\pi)^3} \, \frac{\mathbf{k} \cdot \mathbf{k}'}{(k')^2} \delta_{\rm a}(\mathbf{k}-\mathbf{k}') \delta_{\rm m}(\mathbf{k}').
    \end{align}
    This expression results in three contributions from the various cross-terms. The first is a purely linear term, proportional to the matter power spectrum:
    \begin{equation}\label{eq:P_mm}
      P^{\rm mm}_\parallel(k)\equiv(aHf)^2\frac{P_{\rm mm}(k)}{k^2}.
    \end{equation}
    Additionally, there are two mixed linear-convolution terms, $P^{\rm gmm}_\parallel$ and $P^{\rm emm}_\parallel$, involving the galaxy-matter and electron-matter bispectra, respectively:
    \begin{align}
      P_{\parallel}^{\rm amm}(k) &= \frac{(aHf)^2}{(2\pi)^2} \int \mathrm{d}k' \, \frac{k'}{k} \int_{-1}^1 \mathrm{d}\mu' \, \mu' B_{\rm amm}(k, k', \mu'), \label{eq:P_par_amm}
    \end{align}
    where $\mu' = \hat{\mathbf{k}} \cdot \hat{\mathbf{k}}'$ as before. Here, the bispectrum is defined as
    \begin{equation}
      \langle \delta_{\rm a}(\mathbf{k}_1) \delta_{\rm b}(\mathbf{k}_2) \delta_{\rm c}(\mathbf{k}_3) \rangle = (2\pi)^3 \delta^{\mathcal{D}}(\mathbf{k}_1 + \mathbf{k}_2 + \mathbf{k}_3) B_{\rm abc}(\mathbf{k}_1, \mathbf{k}_2, \mathbf{k}_3).
    \end{equation}
    The final contribution arises from the convolution-convolution term, and can be split into three terms that have the same structure as those found for the transverse modes (Equations \eqref{eq:P_perp_1}, \eqref{eq:P_perp_2}, and \eqref{eq:P_perp_c}):
    \begin{align}
      P^{\rm ge}_{\parallel,1}(k) &= (aHf)^2 \int \frac{\mathrm{d}^3 k'}{(2\pi)^3} \frac{(\mathbf{k} \cdot \mathbf{k}')^2}{k^2 (k')^4} P_{\rm ge}(|\mathbf{k}-\mathbf{k}'|) P_{\rm mm}(k') , \\[0.5em]
      P^{\rm ge}_{\parallel,2}(k) &= (aHf)^2 \int \frac{\mathrm{d}^3 k'}{(2\pi)^3} \frac{1}{|\mathbf{k}-\mathbf{k}'|^2} \frac{k}{ k'} \mu' \left(1-\frac{k'}{k}\mu' \right) P_{\rm gm}(|\mathbf{k}-\mathbf{k}'|)\,P_{\rm em}(k'), \\[0.5em]
      P^{\rm ge}_{\parallel, \rm c}(k) &= (aHf)^2 \int \frac{\mathrm{d}^3k'}{(2\pi)^3} \int \frac{\mathrm{d}^3k''}{(2\pi)^3} \, \left(\frac{(\mathbf{k} \cdot \mathbf{k}') (\mathbf{k} \cdot \mathbf{k}'')}{k^2 (k')^2 (k'')^2}\right) T_{\rm gemm}(\mathbf{k} - \mathbf{k}', -(\mathbf{k} - \mathbf{k}''), \mathbf{k}', -\mathbf{k}'').
    \end{align}
    In spherical coordinates, these components become:
    \begin{align}
        P^{\rm ge}_{\parallel, 1}(k) &= \frac{(aHf)^2}{(2\pi)^2} \int \mathrm{d}k' \,  P_{\rm mm}(k') \int_{-1}^1 \mathrm{d}\mu \, \mu^2 P_{\rm ge}(|\mathbf{k}-\mathbf{k}'|), \label{eq:P_par_1} \\[0.5em]
        P^{\rm ge}_{\parallel, 2}(k) &= \frac{(aHf)^2}{(2\pi)^2} \int \mathrm{d}k' \, P_{\rm gm}(k') \int_{-1}^{1} \mathrm{d}\mu' \, \frac{k k' \mu'}{|\mathbf{k}-\mathbf{k}'|^2} \left(1-\frac{k'}{k}\mu'\right) P_{\rm em}(|\mathbf{k}-\mathbf{k}'|), \label{eq:P_par_2} \\[0.5em]
        P^{\rm ge}_{\parallel, \rm c}(k) &= \frac{(aHf)^2}{(2\pi)^5} \int \mathrm{d}k' \, k' \int \mathrm{d}k'' \, k'' \int_{-1}^1 \mathrm{d}\mu' \, \mu' \int_{-1}^1 \mathrm{d}\mu'' \, \mu'' \int_0^{2\pi} \mathrm{d}\phi \nonumber \\[0.2em]
        &\hspace{40pt}\times T_{\rm gemm}(\mathbf{k} - \mathbf{k}', -(\mathbf{k} - \mathbf{k}''), \mathbf{k}', -\mathbf{k}''). \label{eq:P_par_c}
    \end{align}
    Hence, the total longitudinal momentum power spectrum can be expressed as the sum of all individual contributions:
    \begin{equation}
      P^{\rm ge}_{\parallel}(k) = P^{\rm ge}_{\parallel, 1}(k) + P^{\rm ge}_{\parallel, 2}(k) + P^{\rm ge}_{\parallel, \rm c}(k) + P^{\rm gmm}_{\parallel}(k) + P_{\parallel}^{\rm emm}(k) + P^{\rm mm}_\parallel(k).
    \end{equation}

   \subsection{Halo model for higher-order statistics} \label{ssec:theory.halo_model}
    The halo model provides a useful framework to estimate the connected bispectrum and trispectrum entering Equations \eqref{eq:P_par_amm}, \eqref{eq:P_perp_c}, and \eqref{eq:P_par_c}. This approach decomposes higher-order statistics into contributions from structures involving one or more haloes. An alternative method for estimating higher-order contributions involves the use of simulations (e.g. \cite{Alvarez:2016, Park:2018baryonic}).

    The matter bispectrum can be expressed as a sum over one-, two-, and three-halo contributions via $B = B_{\rm 1h} + B_{\rm 2h} + B_{\rm 3h}$ \cite{1102.0641}. The one-halo term, which dominates on small scales, is given by
    \begin{equation} \label{eq:B_1h}
        B^{\delta_{\rm A} \delta_{\rm B} \delta_{\rm C}}_{\rm 1h}(\mathbf{k}_1, \mathbf{k}_2, \mathbf{k}_3) = \int \mathrm{d}M \, n(M)\,\langle u_{\delta_{\rm A}}(k_1|M) u_{\delta_{\rm B}}(k_2|M) u_{\delta_{\rm C}}(k_3|M) \rangle,
    \end{equation}
    where $n(M)$ is the mass function of the halo, and the $u_{\delta_{i}}(k|M)$ are the halo profiles in Fourier space that correspond to the $\delta_{i}$. On large scales, the three-halo term becomes dominant, which is given by \begin{align}
        B^{\delta_{\rm A} \delta_{\rm B} \delta_{\rm C}}_{\rm 3h}(\mathbf{k}_1, \mathbf{k}_2, \mathbf{k}_3) &= \bU{A}(k_1)\,\bU{B})(k_2)\,\bU{C}(k_3)\,B_{\rm PT}(\mathbf{k}_1, \mathbf{k}_2, \mathbf{k}_3) \label{eq:B_3h},
    \end{align}
    where
    \begin{equation}
      \bU{A}(k)\equiv \int \mathrm{d}M \, n(M) b_h(M) \langle u_{\delta_{\rm A}}(k|M) \rangle,
    \end{equation}
    and $b_h(M)$ is the linear halo bias. At intermediate scales, the two-halo term is the dominant contribution. However, we do not consider this contribution in our analysis since, as we will show, the bispectrum contribution on large and small scales (as given by the expressions above) is subdominant by a large factor.

    In the three-halo term, $B_{\rm PT}$ denotes the tree-level matter bispectrum, which in turn depends on the second-order perturbation theory kernel, $F_2$, and the linear matter power spectrum, $P_{\rm lin}(k)$, via
    \begin{equation}
        B_{\rm PT}(\mathbf{k}_1, \mathbf{k}_2, \mathbf{k}_3) = 2 F_2(\mathbf{k}_1, \mathbf{k}_2) P_{\rm lin}(k_1) P_{\rm lin}(k_2) + (2,3) + (3,1),
    \end{equation}
    where the second-order density kernel, $F_2$, is given by
    \begin{equation}
        F_2(\mathbf{k}_1, \mathbf{k}_2) \equiv \frac{5}{7} + \frac{\mu_{12}}{2} \left(\frac{k_2}{k_1} + \frac{k_1}{k_2}\right) + \frac{2}{7} \mu_{12}^2,
    \end{equation}
    and $\mu_{12} \equiv (\mathbf{k}_1 \cdot \mathbf{k}_2)/(k_1 k_2)$.

    Similarly, the matter trispectrum can be written as $T = T_{\rm 1h} + T_{\rm 2h,31} + T_{\rm 2h, 22} + T_{\rm 3h} + T_{\rm 4h}$. The one-halo term dominates on small-scales and is given by
    \begin{equation}
        T^{\delta_{\rm A} \delta_{\rm B} \delta_{\rm C} \delta_{\rm D}}_{\rm 1h}(\mathbf{k}_1, \mathbf{k}_2, \mathbf{k}_3, \mathbf{k}_4) = \int \mathrm{d}M \, n(M) \left\langle u_{\delta_{\rm A}}(k_1|M) u_{\delta_{\rm B}}(k_2|M) u_{\delta_{\rm C}}(k_3|M) u_{\delta_{\rm D}}(k_4|M) \right\rangle, \label{eq:T_1h}
    \end{equation}
    On the largest scales, the four-halo term becomes dominant and takes the form:
    \begin{align}
        T^{\delta_{\rm A} \delta_{\rm B} \delta_{\rm C} \delta_{\rm D}}_{\rm 4h}(\mathbf{k}_1, \mathbf{k}_2, \mathbf{k}_3, \mathbf{k}_4) &= \bU{A}(k_1)\,\bU{B}(k_2)\,\bU{C}(k_3)\,\bU{D}(k_4)\,T^{\delta_{\rm A} \delta_{\rm B} \delta_{\rm C} \delta_{\rm D}}_{\rm PT}(\mathbf{k}_1, \mathbf{k}_2, \mathbf{k}_3, \mathbf{k}_4).
    \end{align}
    Here, $T_{\rm PT}$ is the tree-level matter trispectrum, which itself can be decomposed as $T_{\rm PT} = T_{1122} + T_{1113}$, where each term depends on perturbative kernels and the linear matter power spectrum. The term $T_{1113}$ is given by
    \begin{equation}
        T_{1113}(\mathbf{k}_1, \mathbf{k}_2, \mathbf{k}_3, \mathbf{k}_4) = 6[F_3(\mathbf{k}_1, \mathbf{k}_2, \mathbf{k}_3) P_{\rm lin}(k_1) P_{\rm lin}(k_2) P_{\rm lin}(k_3) + (234) + (341) + (412)],
    \end{equation}
    with the third-order density kernel $F_3$ defined as
    \begin{equation}
        F_3(\mathbf{k}_1, \mathbf{k}_2, \mathbf{k}_3) \equiv \frac{1}{54}[Q(\mathbf{k}_1, \mathbf{k}_2, \mathbf{k}_3) + Q(\mathbf{k}_2, \mathbf{k}_3, \mathbf{k}_1) + Q(\mathbf{k}_3, \mathbf{k}_1, \mathbf{k}_2)].
    \end{equation}
    The kernel $Q$ is in turn given by
    \begin{equation}
        Q(\mathbf{k}_1, \mathbf{k}_2, \mathbf{k}_3) \equiv 7 \frac{\mathbf{k}_{123} \cdot \mathbf{k}_1}{k_1^2} F_2(\mathbf{k}_2, \mathbf{k}_3) + \left(7 \frac{\mathbf{k}_{123} \cdot \mathbf{k}_{23}}{k_{23}^2} + 2 \frac{k_{123}^2 (\mathbf{k}_{23} \cdot \mathbf{k}_1)}{k_{23}^2 k_1^2}\right) G_2(\mathbf{k}_2, \mathbf{k}_3),
    \end{equation}
    and involves the second-order density kernel, $F_2$, and the second-order velocity kernel, $G_2$,
    \begin{equation}
        G_2(\mathbf{k}_1, \mathbf{k}_2) \equiv \frac{3}{7} + \frac{\mu_{12}}{2} \left(\frac{k_1}{k_2} + \frac{k_2}{k_1}\right) + \frac{4}{7}\mu_{12}^2.
    \end{equation}
    The term $T_{1122}$ is given by
    \begin{align}
        T_{1122}(\mathbf{k}_1, \mathbf{k}_2, \mathbf{k}_3, \mathbf{k}_4) = 4[ &Q_{12,3} + Q_{12,4} + Q_{13,2} + Q_{13,4} + Q_{14,2} + Q_{14,3} \nonumber \\
        + &Q_{23,1} + Q_{23,4} + Q_{24,1} + Q_{24,3} + Q_{34,1} + Q_{34,2}],
    \end{align}
    where we have defined
    \begin{equation}
        Q_{\rm ab,c} \equiv F_2(\mathbf{k}_{\rm a} + \mathbf{k}_{\rm c}, -\mathbf{k}_{\rm a}) F_2(\mathbf{k}_{\rm a} + \mathbf{k}_{\rm c}, \mathbf{k}_{\rm b}) P_{\rm lin}(|\mathbf{k}_{\rm a} + \mathbf{k}_{\rm c}|) P_{\rm lin}(k_{\rm a}) P_{\rm lin}(k_{\rm b}).
    \end{equation}

   Using this formalism, one can show that the one-halo contribution to the trispectrum for the transverse power spectrum vanishes in Equation \eqref{eq:P_perp_c}, as also shown in \cite{Ma:2001}. This results from the fact that the halo profiles $u_i(k|M)$ in Equation \eqref{eq:T_1h} depend only on the magnitude of their arguments. Consequently, the angular integration in Equation \eqref{eq:P_perp_c} evaluates to zero due to the presence of the $\cos\phi$ factor. Therefore, we will only consider the contribution from the four-halo term in order to estimate the impact of connected four-point correlations in the kSZ observables for the transverse mode.

  \subsection{Covariance} \label{ssec:theory.cov}
   We calculate the covariance for the angular power spectra using the Knox formula \cite{astro-ph/9504054},
   \begin{equation}
       \mathrm{Cov}(C_{\ell}^{\rm ge}, C_{\ell'}^{\rm ge}) = \frac{\delta_{\ell \ell'}}{(2\ell+1) f_{\rm sky}} \left(C_{\ell}^{\rm gg} C_{\ell}^{\rm ee} + (C_{\ell}^{\rm ge})^2\right),
   \end{equation}
   where $f_{\rm sky}$ is the fraction of the sky covered by the our observations. Here we will assume $f_{\rm sky}=0.5$, roughly corresponding to the overlapping sky coverage of future southern-hemisphere experiments, such as LSST and SO \cite{LSST:2008, LSST:2018, SO:2018}. Hence, in order to compute the covariance, we must also calculate the auto-correlations, $C_{\ell}^{\rm gg}$ and $C_{\ell}^{\rm ee}$. For $C_{\ell}^{\rm gg}$, we have:
   \begin{align}\label{eq:clgg}
     C_{\ell}^{\rm gg} = \widetilde{C}_{\ell}^{\rm gg} + N_{\ell}^{\rm gg},
   \end{align}
   where $\widetilde{C}_{\ell}^{\rm gg}$ denotes the signal contribution and $N_{\ell}^{\rm gg}$ denotes the noise angular power spectrum. We approximate the latter by $N_{\ell}^{\rm gg} = \sigma_v^2 / \bar{n}$, where $\sigma_v^2$ is the velocity dispersion of galaxies and $\bar{n}$ is the number density of galaxies. For $C_{\ell}^{\rm ee}$, we have:
   \begin{equation}
       C_{\ell}^{\rm ee} = C_{\ell}^{\rm CMB} + C_{\ell}^{\rm sec} + N_{\ell}^{\rm ee},
   \end{equation}
   where $C_{\ell}^{\rm CMB}$ is the angular power spectrum of the CMB, $C_{\ell}^{\rm sec}$ denotes the contribution from secondary anisotropies, and $N_{\ell}^{\rm ee}$ is the noise angular power spectrum. The angular power spectrum for the secondary anisotropies can in turn be decomposed as
   \begin{equation}
       C_{\ell}^{\rm sec} = C_{\ell}^{\rm kSZ} + C_{\ell}^{\rm tSZ} + C_{\ell}^{\rm CIB},
   \end{equation}
   where $C_{\ell}^{\rm kSZ}$, $C_{\ell}^{\rm tSZ}$, and $C_{\ell}^{\rm CIB}$ are the angular power spectra for the kSZ effect, tSZ effect, and cosmic infrared background (CIB), respectively. We use the best fit model for $C_\ell^{\rm sec}$ at 90 GHz used in the latest analysis of the  ACT collaboration \cite{ACT:2025}.
   
   The perpendicular and parallel components of $\widetilde{C}_{\ell}^{\rm gg}$ and $\widetilde{C}_{\ell}^{\rm ee} \equiv C_{\ell}^{\rm kSZ}$ can be calculated in the same way as those of $C_{\ell}^{\rm ge}$ in Sections \ref{ssec:theory.perp} and \ref{ssec:theory.par}, respectively. We can then compute the signal-to-noise ratio, $S/N$, of a particular CMB survey for the different contributions to the angular power spectrum via
   \begin{equation}
       \left(\frac{S}{N}\right)_i = \left[\sum_{\ell} \frac{\left(C_{\ell, i}^{\rm ge}\right)^2}{\mathrm{Var}\left(C_{\ell}^{\rm ge}\right)}\right]^{1/2},
    \end{equation}
    where $C_{\ell,i}^{\rm ge}$ is the angular power spectrum of the particular contribution under consideration (e.g. the parallel momentum mode).

 \subsection{Implementation} \label{ssec:theory.implementation}
  To implement the theoretical framework numerically, we use the Core Cosmology Library (CCL) \cite{CCL} to compute halo model quantities under the following parametrisations. We adopt a halo mass definition corresponding to an overdensity threshold of $\Delta = 200$ with respect to the critical density, $\overline{\rho}_{\rm c}$. We use the concentration-mass relation, $c(M)$, from \cite{Duffy:2008}, the halo mass function, $n(M)$, from \cite{Tinker:2008}, and the halo bias, $b_h(M)$, from \cite{Tinker:2010}. Throughout this analysis, we assume a fiducial cosmology consistent with {\sl Planck} results, characterised by the parameter values $\{\Omega_{\mathrm{c}}, \Omega_{\mathrm{b}}, h, n_{\mathrm{s}}, \sigma_8, \Sigma m_{\nu}\} =$ $\{0.2607$, $0.04897$, $0.6766$, $0.9665$, $0.8102, 0.06\}$ \cite{Planck:2018}.
   
  To model the relevant overdensity fields, we adopt the following prescriptions. The matter overdensity, $\delta_{\rm m}$, is described using the NFW profile \cite{Navarro:1997}. The galaxy overdensity, $\delta_{\rm g}$, is modelled within the Halo Occupation Distribution (HOD) framework \cite{Zheng:2004, Ando:2017, LSST:2019}, which we will detail below. The electron density profile, $\rho_{\rm e}$, is computed using the hydrostatic equilibrium model used in \cite{Ferreira:2023, LaPosta:2025}. This approach is motivated by previous efforts such as baryonification and the HMx framework \cite{Schneider:2015, Mead:2020}. 
  
  To initialise the electron density profile, we adopt the following fiducial baryonic parameters: $\{\log_{10} M_{\rm c}, \beta, \eta_{\rm b}, A_*\} = \{14.0, 0.6, 0.5, 0.03\}$. These values are consistent with current constraints from hydrodynamical analyses \cite{LaPosta:2025}. Here, $M_{\rm c}$ is the mass at which half the gas has been ejected from the halo via baryonic feedback, $\beta$ describes the dependence of $M_{\rm c}$ on halo mass, $\eta_{\rm b}$ characterises the radius to which ejected gas is expelled, and $A_*$ describes the stellar abundance. Finally, we normalise the electron density profile to yield the electron overdensity, $\delta_{\rm e} = \rho_{\rm e}/\overline{\rho}_{\rm e} -1$.
   
  Equipped with these models for the various overdensity fields, we compute all power spectrum terms that appear in the integrals in Equations \eqref{eq:P_perp_1} and \eqref{eq:P_perp_2} for the transverse mode, and in Equations \eqref{eq:P_par_1} and \eqref{eq:P_par_2} for the longitudinal mode. We then propagate the 3D power spectra into angular power spectra by constructing tracer objects using CCL. To ensure that we exhaust all the information available to a given CMB experiment, we consider angular multipoles up to $\ell_{\rm max} = 10,000$. 
  
  As noted earlier, we model the galaxy overdensity using the HOD framework, adopting the parameter values from the second row of Table 4 in \cite{Zhou:2020}. We focus on a luminous red galaxy (LRG) sample, consistent with previous kSZ studies, where LRGs have been the primary observational targets \cite{Hadzhiyska:2024}. To match the density of photometric LRG samples, we assume a number density of 150 galaxies per steradian. We use the second redshift bin from \cite{Zhou:2020} to model the redshift distribution of sources. The HOD model employed here uses the following parametrisations for the mean number of central and satellite galaxies, denoted by $\bar{N}_{\rm c}$ and $\bar{N}_{\rm s}$, respectively:
   \begin{equation}
      \bar{N}_{\rm c}(M,a) = \frac{1}{2} \left[1 + \mathrm{erf}\left(\frac{\log_{10}(M/M_{\rm min})}{\sigma_{\ln M}}\right) \right],
   \end{equation}
  \begin{equation}
     \bar{N}_{\rm s}(M,a) = \Theta(M-M_0) \left(\frac{M-M_0}{M_1}\right)^{\alpha}.
  \end{equation}
  Here, $M_0$ represents the minimum halo mass required to host satellite galaxies, $M_1$ is the characteristic mass scale at which haloes host one satellite galaxy on average, and $M_{\rm min}$ is the threshold mass for central galaxy occupation. The parametrisation for the mean profile is then given by
  \begin{equation}
      \langle n_{\rm g}(r) | M\rangle = \bar{N}_{\rm c}(M) \left[f_{\rm c} + \bar{N}_{\rm s}(M) u_{\rm sat}(r | M)\right],
  \end{equation}
  where $f_{\rm c}$ is the observed fraction of central galaxies and $u_{\rm sat}(r | M)$ is the distribution of satellite galaxies as a function of distance to the halo centre, which we assume to be given by the Navarro-Frenk-White profile \cite{Navarro:1997}

  The fiducial HOD model of \cite{Zhou:2020} used here assumes $f_{\rm c}=1$, and the resulting fraction of satellites in the sample is $f_{\rm sat}=0.14$. To study the impact of satellites on the kSZ signal, we vary the value of $f_{\rm c}$, simultaneously varying $M_{\rm min}$ and fixing $M_0=M_{\rm min}$ to ensure that the large-scale galaxy bias $b_{\rm g}$ remains unchanged with respect to the fiducial model. 

  The significance of the kSZ signal and its individual contributions depends critically on the noise properties of the CMB map used, which are characterised by the noise power spectrum $N_\ell^{\rm ee}$. We consider three different scenarios for the noise power spectra. Firstly, we adopt noise levels consistent with those anticipated for SO \cite{SO:2018}. In particular, we use the official temperature noise curves provided in \cite{SO:2018} for the baseline noise level of the SO Large Aperture Telescope (LAT), assuming a $2.1\,{\rm arcmin}$ FWHM Gaussian beam. Secondly, we consider the forecasted noise power spectrum for the CMB-S4 experiment. For this case, we assume the noise parameters specified for the 90 GHz channel in Table 1 of \cite{S4:2024}. We note that CMB-S4 has recently been discontinued. We nevertheless present forecasts for this configuration as a proxy for a future experiment succeeding SO. Finally, we consider an idealised case, in which instrumental noise is neglected entirely, corresponding to the cosmic variance limit (CVL).

  The trispectrum $T(\mathbf{k}, \mathbf{k}', \mathbf{k}'')$ appearing in Equations \eqref{eq:P_perp_c} and \eqref{eq:P_par_c} is computed via direct numerical integration over five variables: the magnitudes $k'$ and $k''$, the angular coordinates $\mu_1 = \cos\theta_1$ and $\mu_2 = \cos\theta_2$, and the azimuthal angle $\phi$ between $\mathbf{k}'$ and $\mathbf{k}''$. The integration is performed on regular grids, with logarithmically spaced points for each wavenumber and uniformly spaced points for the angular and azimuthal coordinates. At each grid point, we evaluate the trispectrum function directly, and the five-dimensional integral is computed using successive one-dimensional trapezoidal integrations over each variable. Convergence tests with finer grids were carried out to confirm the stability of the results.

 \section{Results} \label{sec:results}
   In this section, we quantify the impact of the different transverse and longitudinal terms that contribute to the kSZ-galaxy cross-power spectrum. We will focus first on the dominant transverse modes, and then examine the longitudinal contributions. Using the specific implementation described in Section \ref{ssec:theory.implementation}, we provide estimates for the significance of the different contributions in Section \ref{ssec:results.sensitivity} (Table \ref{tab:S/N}).

  \subsection{Contributions to the transverse power spectrum} \label{ssec:results.transverse}
\begin{figure*}[h!]
       \centering
       \includegraphics[width=0.485\linewidth]{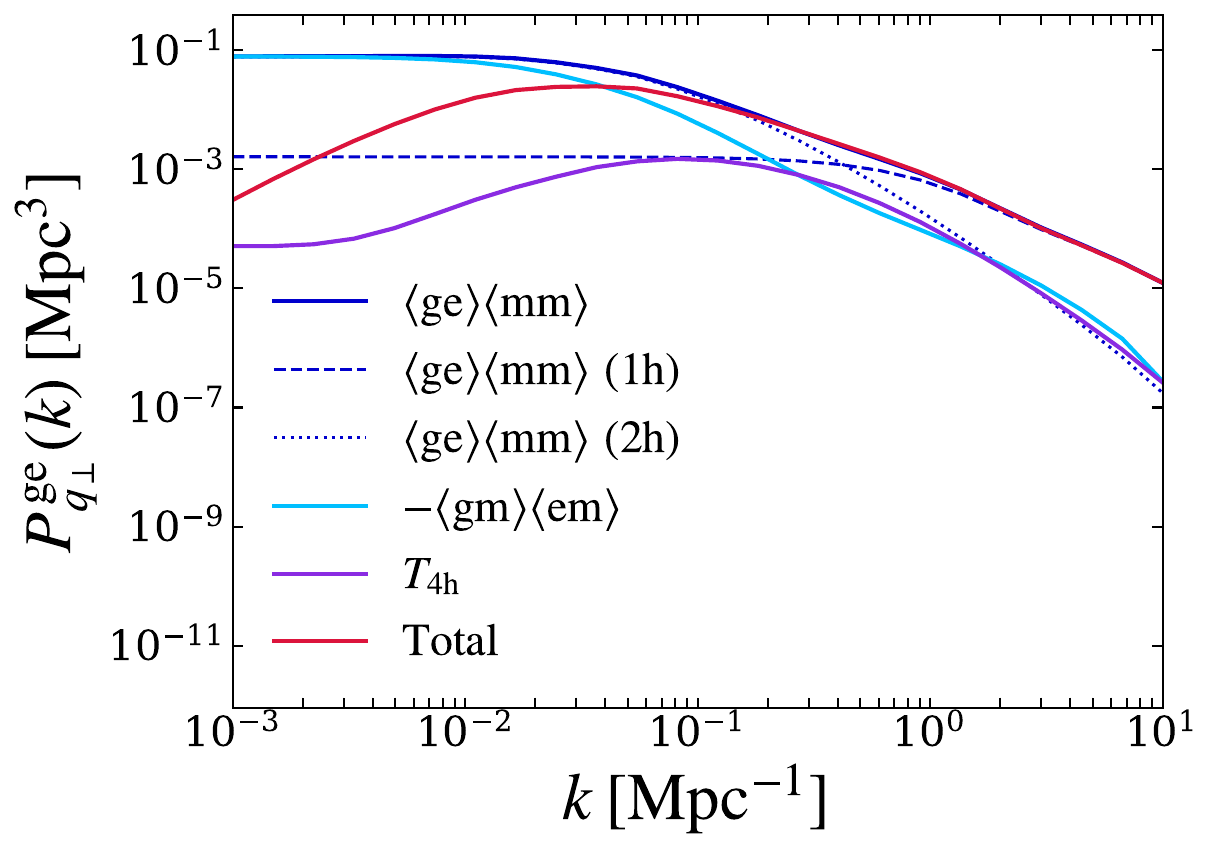}
       \includegraphics[width=0.49\linewidth]{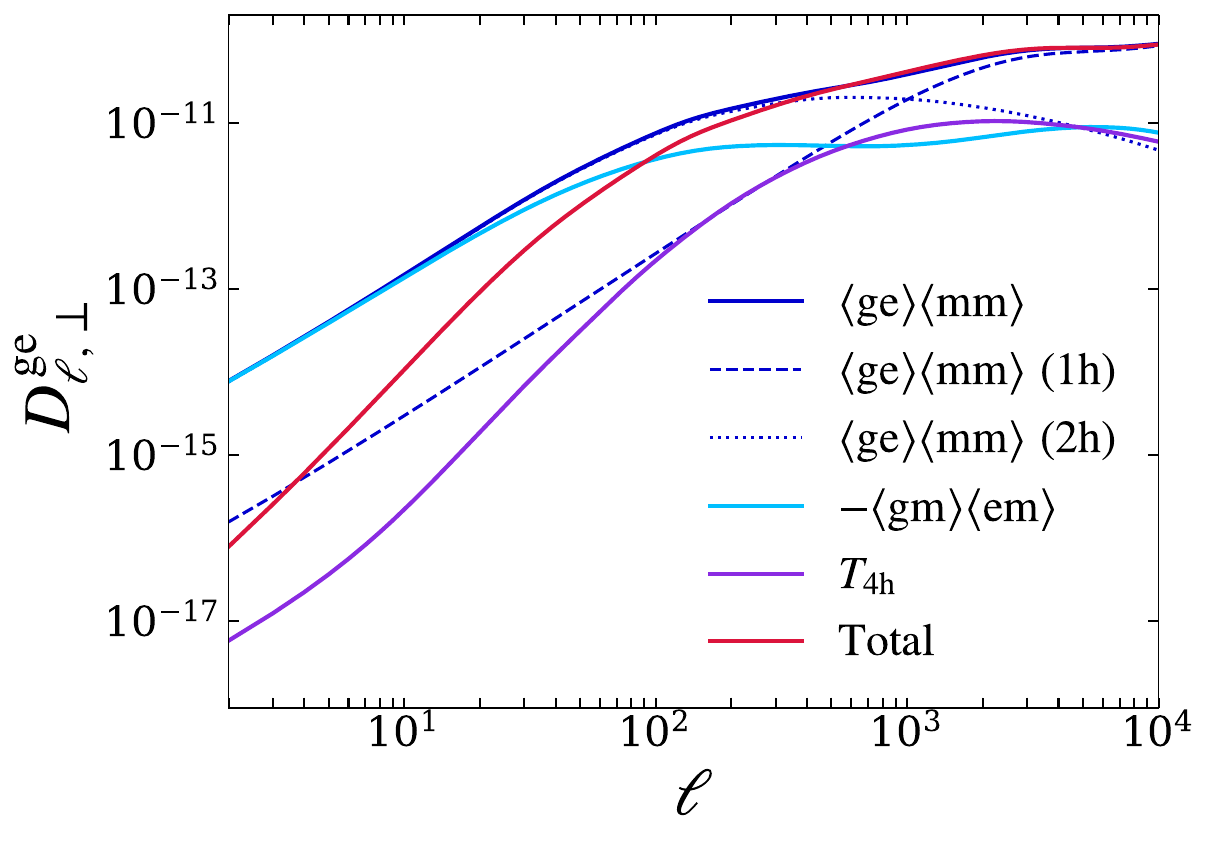}
       \caption{The contributions to the 3D power spectrum (left) and angular power spectrum (right) for the perpendicular component of the momentum field. The individual contributions from $\rm \langle ge \rangle \langle mm \rangle$ (dark blue), $\rm \langle gm \rangle \langle em \rangle$ (light blue), and the connected non-Gaussian term (purple) are shown, with their total displayed in red. The dashed and dotted lines denote the $\rm \langle ge \rangle \langle mm \rangle$ term for the one-halo and two-halo contributions, respectively, in the electron-galaxy cross-correlation.}
       \label{fig:transverse}
   \end{figure*}

   In Figure \ref{fig:transverse}, we present the contributions to both the 3D and angular power spectra for the transverse mode of the momentum field. The angular power spectrum is expressed in terms of $D_{\ell} \equiv [\ell(\ell+1)/(2\pi)] C_{\ell}$. The individual contributions from Equations \eqref{eq:P_perp_1} and \eqref{eq:P_perp_2} are shown in dark blue and light blue, respectively, with their total shown in red. The dashed and dotted lines denote the one-halo and two-halo contributions to the galaxy-electron cross-correlation, $P_{\rm ge}(k)$, in the $\rm \langle ge \rangle \langle mm \rangle$ term, respectively. 
   
   We find that the contribution from $P_{\perp, \, 1}$ (given by Equation \eqref{eq:P_perp_1}) dominates the power spectrum on small scales $(k \gtrsim 0.1 \, \mathrm{Mpc}^{-1})$, while the contribution from $P_{\perp, \, 2}$ (given by Equation \eqref{eq:P_perp_2}) becomes significant only on large scales $(k \lesssim 0.1 \, \mathrm{Mpc}^{-1})$. Interestingly, $P_{\perp, \, 1}$ and $P_{\perp, \, 2}$ have equal amplitudes but opposite signs at small $k$, resulting in a cancellation of the total 3D power spectrum at $k=0$. 
   
   The purple line in Figure \ref{fig:transverse} shows the contribution from the connected trispectrum term, as defined in Equation \eqref{eq:P_perp_c}. We find that this term is subdominant on large scales but becomes increasingly significant at smaller scales, having a contribution comparable in magnitude to the cross-term $P_{\perp,2}$ on scales $k\gtrsim 0.3\,{\rm Mpc}^{-1}$. We note that the trispectrum contribution in the left panel of Figure \ref{fig:transverse} flattens towards low $k$, in contrast to the results of \cite{Park:2016}, where this term scales as $k^2$ in the same regime. This discrepancy arises from our assumption that all velocity fields are linearly related to the density field through the continuity equation. As a result, we caution that our results may not be reliable at low $k$. Moreover, in our analysis, the projected galaxy momentum field does not represent a true momentum field, since the velocity entering Equation \eqref{eq:pi_g} corresponds to a reconstructed velocity rather than the true velocity. Therefore, the perturbative approach should not be expected to hold in the low-$k$ limit, as doing so would implicitly treat the reconstructed velocity as exact.

   The dashed line in Figure \ref{fig:transverse} shows the small-scale approximation described in Section \ref{sssec:theory.perp.hik} (Equation \ref{eq:ptrans_hik}), having retained only the $P_{\perp,1}$ contribution, and discarding the two-halo term of $P_{\rm ge}$. This one-halo term captures intra-halo clustering, describing correlations between galaxies and electrons within the same halo or, in the case of a sample of central galaxies, the distribution of gas around the host haloes. As expected, the one-halo term dominates on small scales ($k\gtrsim1\,{\rm Mpc}^{-1}$), effectively recovering the total power spectrum in this regime, while the two-halo term becomes increasingly significant on intermediate and large scales. These scale-dependent features are also reflected in the angular power spectrum, shown in the right-hand panel of Figure \ref{fig:transverse}, demonstrating consistency between the 3D and projected statistics, which is enabled by the well-localised redshift bin assumed here.

  \subsection{Contributions to the longitudinal power spectrum} \label{ssec:results.longitudinal}

   \begin{figure*}[h!]
      \centering
      \includegraphics[width=0.48\linewidth]{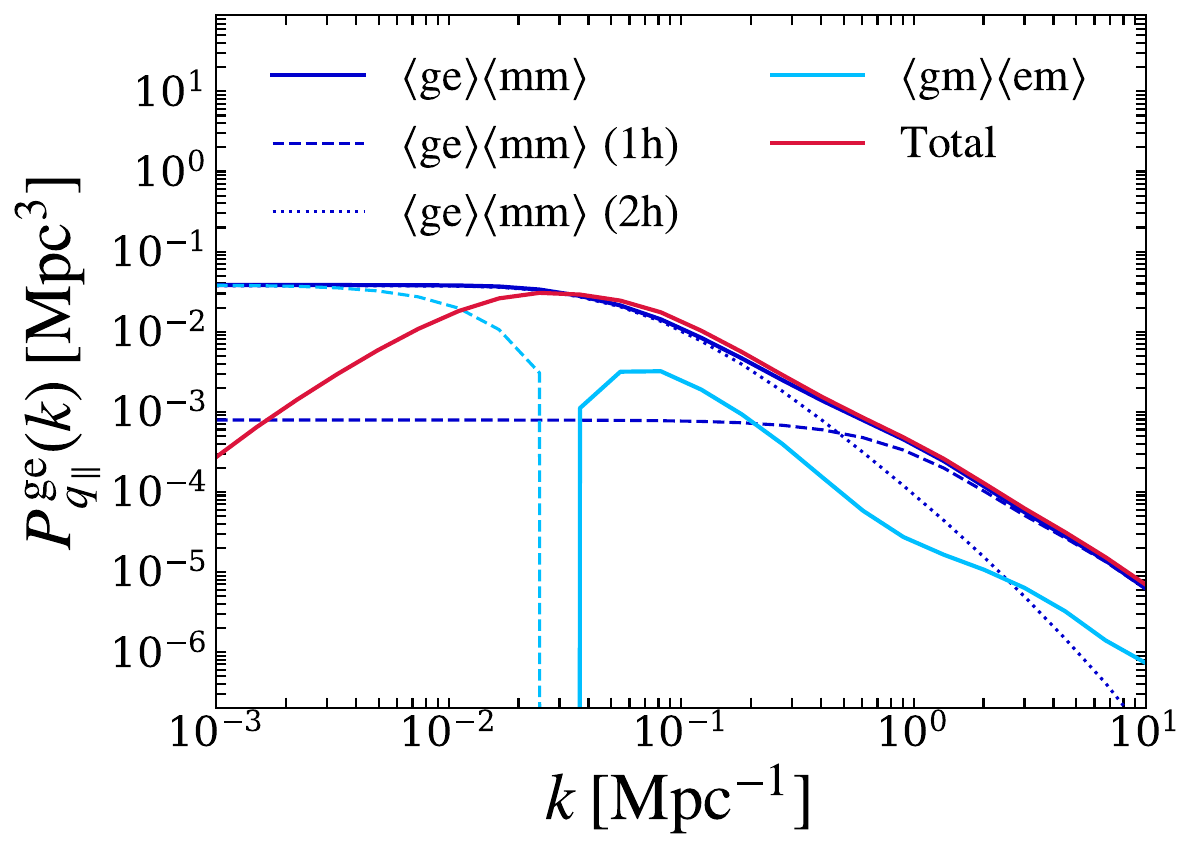}
      \includegraphics[width=0.49\linewidth]{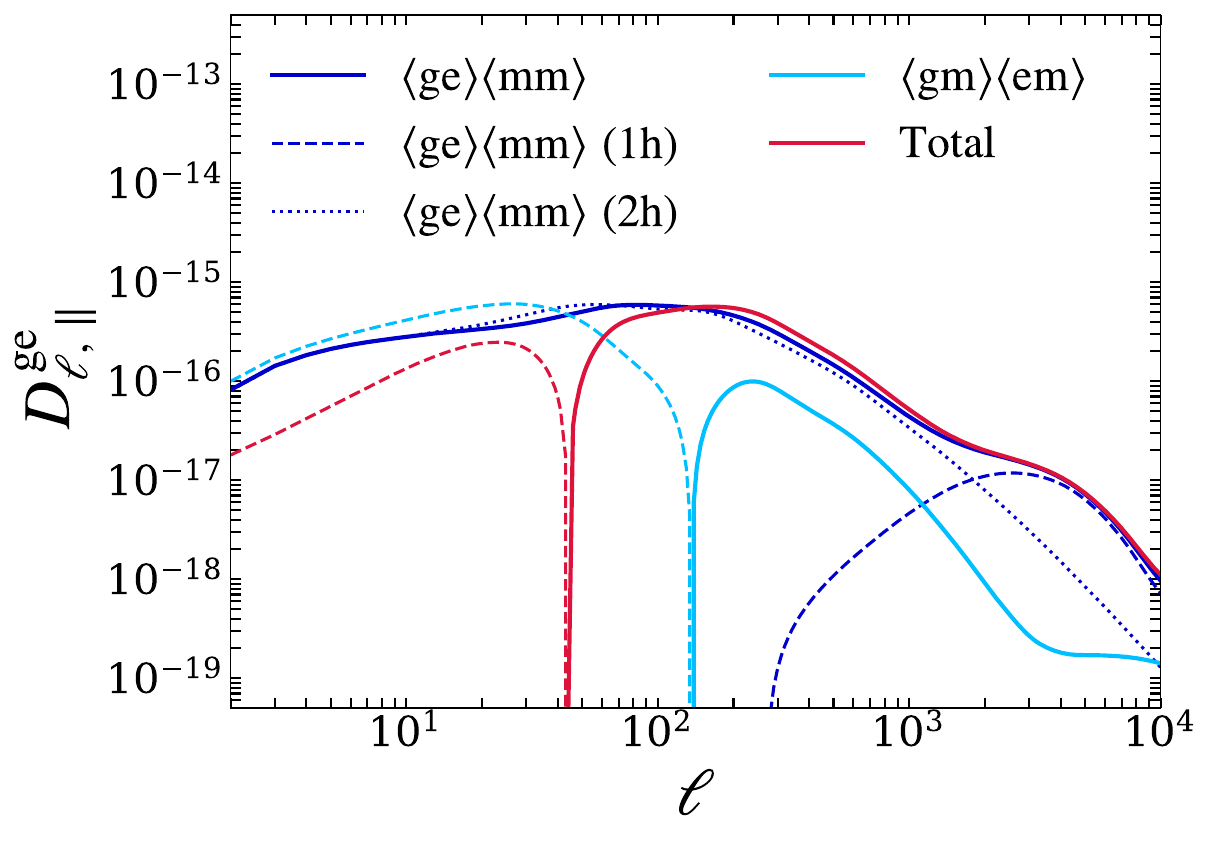}
      \caption{The contributions to the 3D power spectrum (left) and angular power spectrum (right) for the parallel component of the momentum field. The contributions from the $\rm \langle ge \rangle \langle mm \rangle$ and $\rm \langle gm \rangle \langle em \rangle$ terms are shown in dark blue and light blue, respectively, with their total shown in red. The dashed and dotted lines denote the $\rm \langle ge \rangle \langle mm \rangle$ term for the one-halo and two-halo contributions, respectively, in the electron-galaxy cross-correlation.}
      \label{fig:longitudinal}
   \end{figure*}
   
   \begin{figure*}[h!]
      \centering
      \includegraphics[width=0.48\linewidth]{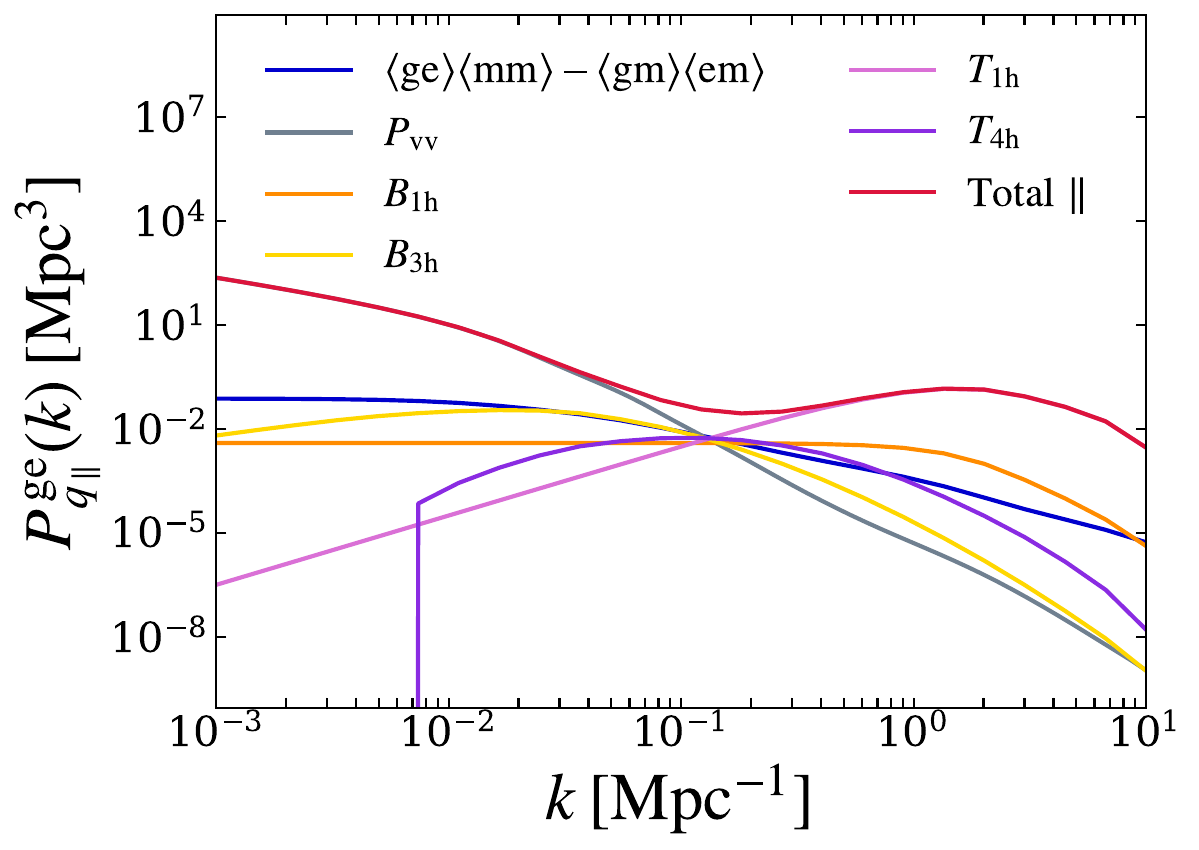}
      \includegraphics[width=0.49\linewidth]{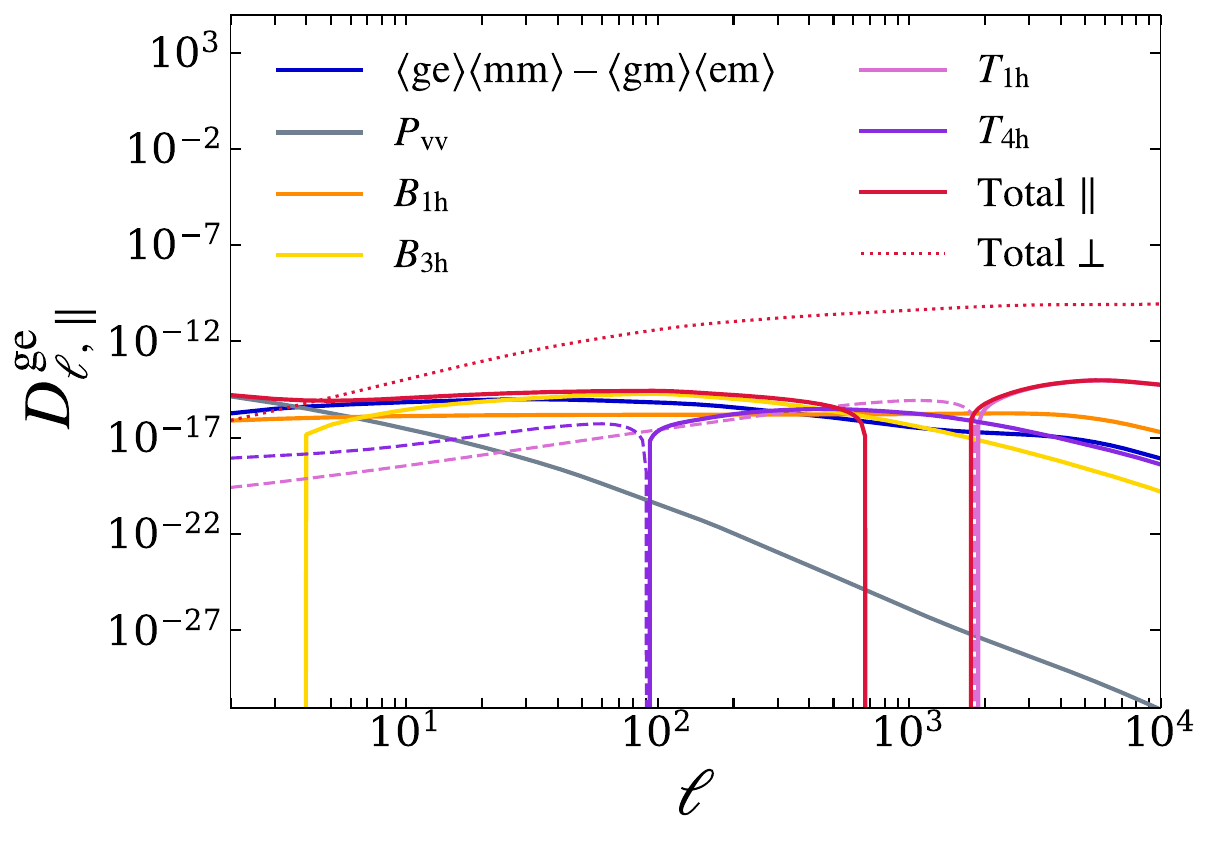}
      \caption{The contributions to the 3D power spectrum (left) and angular power spectrum (right) for the parallel component of the momentum field, including the bispectrum and trispectrum contributions from Equations \eqref{eq:P_par_amm} and \eqref{eq:P_par_c}, respectively. The contributions from $\rm \langle ge \rangle \langle mm \rangle - \rm \langle gm \rangle \langle em \rangle$ (dark blue), one-halo bispectrum (orange), three-halo bispectrum (yellow), one-halo trispectrum (pink), and four-halo trispectrum (purple) are shown, with their total displayed in red. The contribution from the purely linear term in Equation \eqref{eq:P_mm}, $P_{\rm vv}$, is shown in grey. We include the total power of the transverse mode in dotted red. Negative contributions are shown as dashed lines. Note that the region below $k \lesssim 0.01 \, \rm Mpc^{-1}$ for the $T_{\rm 4h}$ contribution to $P_{q_\parallel}^{\rm ge}(k)$ has been masked due to numerical precision issues.}
      \label{fig:longitudinal_with_cng}
   \end{figure*}
   
   In Figure \ref{fig:longitudinal}, we present the contributions to the 3D and angular power spectra for the longitudinal mode of the momentum field. The individual contributions from Equations \eqref{eq:P_par_1} and \eqref{eq:P_par_2} are shown in dark blue and light blue, respectively, with their total shown in red. The dashed and dotted lines indicate the one-halo and two-halo contributions to the galaxy-electron cross-correlation, $P_{\rm ge}(k)$, in the $\rm \langle ge \rangle \langle mm \rangle$ term, respectively. As in the transverse case, we find that $P_{\parallel,1}$ alone (given by Equation \eqref{eq:P_par_1}) is sufficient to describe the power spectrum on small scales. In turn, the longitudinal mode $P_{\parallel,2}$
   becomes relevant at larger scales $(k \lesssim 0.01 \, \mathrm{Mpc}^{-1})$. As before, the one-halo term captures the behaviour of $P_{\parallel,1}$ on the smallest scales $(k \gtrsim 1 \, \mathrm{Mpc}^{-1})$.
   
   These qualitative results are also evident in the angular power spectrum, shown in the right-hand panel of Figure \ref{fig:longitudinal}. However, it is important to note the significant drop in amplitude for the longitudinal angular power spectrum with respect to its transverse counterpart in Figure \ref{fig:transverse}, with a decrease of several orders of magnitude across all scales. This reduction is due to the highly oscillatory nature of the complex exponential in Equation \eqref{eq:pi_par_FT}, as described in Section \ref{ssec:theory.par}. We could have also expected this by examining the expression for the angular power spectrum in Equation \eqref{eq:C_ell_par}: the derivative of the Bessel function can be expressed as  $j_\ell'=(\ell/x)j_\ell(x)-j_{\ell+1}(x)$, which is heavily suppressed in the Limber regime ($\ell\sim x$). Consequently, we can expect the longitudinal contributions to be negligible.

   In Figure \ref{fig:longitudinal_with_cng}, we present an updated version of Figure \ref{fig:longitudinal}, now including the contributions from the velocity power spectrum (Equation \ref{eq:P_mm}), as well as the the bispectrum and trispectrum terms, described by Equations \eqref{eq:P_par_amm} and \eqref{eq:P_par_c}, respectively. We find that the amplitudes of these terms are non-negligible, and in fact can surpass that of $P_{\parallel,1}+P_{\parallel,2}$. In particular, we find that the one-halo trispectrum contribution dominates the overall signal on small scales. This is contrary to the results reported by \cite{Ma:2001} in the context of the kSZ auto-spectrum, who find that this contribution decays as a steep power of $k$. We instead find that the contribution rises as $\propto k^2$, eventually dominating the signal\footnote{This behaviour can be shown analytically by examining Equation \eqref{eq:P_par_c}. The one-halo trispectrum depends only on the moduli of the wave-vectors $k'$, $k''$, $|{\bf k}-{\bf k}'|$, and $|{\bf k}-{\bf k}''|$. This gives it a specific dependence on the angle cosines $\mu'$ and $\mu''$, such that the integral in Equation \eqref{eq:P_par_c} vanishes in the limit $k\rightarrow 0$. This is also the case for its first derivative with respect to $k$, but not for the second derivative, thus explaining the $\propto k^2$ behaviour. We thank Boris Bolliet for suggesting the exploration of this analytical limit.}. However, since the longitudinal mode remains subdominant compared to the transverse mode, the impact of these higher-order contributions on the overall signal is still negligible.

  \subsection{Sensitivity of CMB surveys} \label{ssec:results.sensitivity}
   In this section, we identify the components of the kSZ-galaxy angular power spectrum that future CMB surveys will have the capability to detect. We calculate the signal-to-noise ratio, $S/N$, for each contribution, considering the three noise level scenarios detailed in Section \ref{ssec:theory.cov}. Table \ref{tab:S/N} summarises the $S/N$ for the individual terms in the total angular power spectrum. As could be expected, we observe a consistent increase in $S/N$ with improving experimental sensitivity, from SO to CMB-S4 and ultimately to the cosmic variance limit.

   \begin{table}[h!]
      \centering
      \begin{tabular}{llccc}
        \hline
        Effect & Term & $S/N_{\rm SO}$ & $S/N_{\rm S4}$ & $S/N_{\rm CVL}$ \\
        \hline
        \hline
        Total kSZ & $C_{\ell, \perp, \, \rm tot}^{\rm ge}$ & 34.2 & 49.1 & 89.0 \\
        Cross-term $\rm \langle gm\rangle\langle em\rangle$ & $C_{\ell, \perp, \, 2}^{\rm ge}$ & 4.18 & 5.81 & 9.35 \\
        Satellite contribution & $C_{\ell, \perp,\, 1}^{\rm ge, \, \rm (cen+sats)}$ & 5.10 & 12.1 & 37.7 \\
        Two-halo contribution & $C_{\ell, \perp,\, 1}^{\rm ge, \, \rm (2h)}$ & 6.33 & 7.55 & 8.90 \\
        Connected trispectrum & $C_{\ell, \perp, \, \rm c}^{\rm ge}$ & 4.94 & 6.35 & 8.57 \\
        Total longitudinal component & $C_{\ell, \parallel, \, \rm tot}^{\rm ge}$ & $3.1 \times 10^{-3}$ & $4.6 \times 10^{-3}$ & $7.3 \times 10^{-3}$ \\
        \hline
      \end{tabular}
      \caption{The signal-to-noise ratio, $S/N$, for the different contributions to the kSZ signal for three different CMB noise levels: the SO baseline, CMB-S4, and the cosmic variance limit (CVL). Here, the CVL refers to an idealised case with no instrumental noise. We use $\ell_{\rm max} = 10,000$ and the fiducial values of the baryonic parameters.}
      \label{tab:S/N}
   \end{table}

   Our results highlight the significance of the second term, $C_{\ell,\perp, \, 2}^{\rm ge}$, which describes the convolution between the galaxy-matter and electron-matter cross-correlations. We forecast a detection significance of $\sim4\sigma$ for this term with SO, growing to $\sim6\sigma$ for CMB-S4. Moreover, the higher-order connected non-Gaussian contribution yields a notable $S/N$ of $\sim5\sigma$, emphasising the necessity of including this term in the modelling of the kSZ-galaxy angular power spectrum. Additionally, we find that both the contribution from satellite galaxies and the two-halo term yield substantial $S/N$ values (5$\sigma$ and 6$\sigma$ respectively for SO, growing to 12 and 8$\sigma$ for CMB-S4), and therefore cannot be neglected in kSZ analyses. However, it is important to note that the halo model typically underestimates the amplitude of the power spectrum in the transition regime between the one-halo and two-halo scales. This is typically caused by an underestimation of the two-halo contribution on intermediate scales due to the failure of the standard halo model to account for non-linear halo bias \cite{2011.08858,2508.10902}. As a result, our calculated signal-to-noise ratio, $S/N$, for the two-halo term is likely a conservative underestimate of its true value.

   As could be expected from Figure \ref{fig:longitudinal}, even the dominant contribution to the angular power spectrum from longitudinal modes has a negligible signal-to-noise ratio. Thus, even with the enhanced sensitivity of future experiments, the longitudinal contributions remain undetectable at any meaningful significance and can thus be safely excluded from theoretical modelling.
   
   These results highlight the importance of modelling the full correlation between matter density and velocity fields. This correlation is captured by both the $P_{\perp,2}$ cross-term and the connected trispectrum which, as we have shown, have a non-negligible effect on the kSZ-galaxy power spectrum. The sensitivity of current kSZ measurements is likely low enough to allow for a simplified modelling, based on the small-scale approximation of Equation \eqref{eq:ptrans_hik}, and dependent only on the galaxy-electron power spectrum, or even just the gas density profile. This will likely not be the case for future experiments. On the one hand, analytical models of the kSZ signal should incorporate the various additional terms discussed here, describing the large-scale correlations between densities and velocity. On the other hand, simulation-based emulators should be based on estimates of the full correlation between the galaxy and electron momentum fields, and not simply on the gas profiles measured on simulations. 

   It is important to note that our results regarding the total $S/N$ of the kSZ signal are subject to various assumptions made in our calculation. Firstly, we assume a photometric galaxy sample with a density of 150 galaxies per steradian. This introduces significant uncertainty in redshift estimates, which propagates into the reconstructed peculiar velocities. However, we have assumed perfect velocity reconstruction, thereby underestimating the true uncertainty. The main impact of imperfect reconstruction would be a reduction in the amplitude of the measured signal, proportional to the correlation coefficient between the true and reconstructed velocities -- a limitation we will explore in Section \ref{sssec:results.sensitivity.velrec}. On the other hand, we have considered only a single redshift bin, whereas current and future analyses target multiple galaxy samples at different redshifts. Nevertheless, our findings remain robust, as our primary goal is to assess the relative significance of the various contributions relative to the total kSZ signal.

    \begin{figure}
        \centering
        \includegraphics[width=0.6\linewidth]{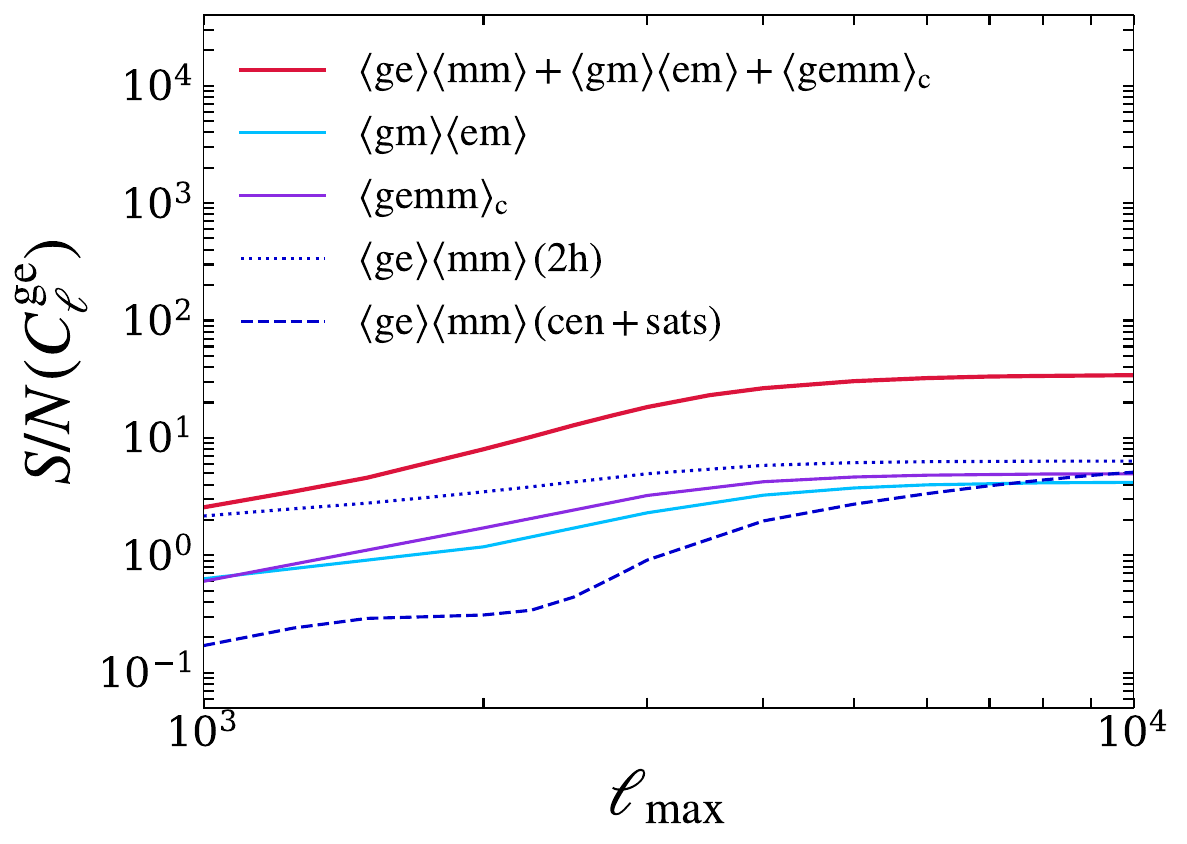}
        \caption{The signal-to-noise ratio, $S/N$, as a function of the angular multipole cut-off, $\ell_{\rm max}$. Here, we fix the baryonic parameters to the fiducial values and we assume the baseline noise level of the Simons Observatory. We also show the $S/N$ of the relevant terms, as listed in Table \ref{tab:S/N}. Specifically, the $\rm \langle gm \rangle \langle em \rangle$ cross-term is shown in light blue, the connected trispectrum term in purple, the two-halo contribution to the $\rm \langle ge \rangle \langle mm \rangle$ term in dotted dark blue, and the $\rm \langle ge \rangle \langle mm \rangle$ term including satellites in the HOD model in dashed dark blue. Here, we compare the satellites case with $f_{\rm sat} = 1.0$  to the fiducial case where $f_{\rm sat} = 0.14$.}
        \label{fig:angular_cutoff}
    \end{figure}
   It is interesting to study the angular scales that contribute most significantly to the kSZ signal explored here. Figure \ref{fig:angular_cutoff} examines the impact of the angular multipole cut-off, $\ell_{\rm max}$, on the signal-to-noise ratio, $S/N$, assuming the SO baseline noise level. We find that the multipole range $2000\lesssim\ell_{\rm max}\lesssim7000$ is critical in order to capture the bulk of the kSZ signal, with the total $S/N$ increasing by a factor $\sim4$ within that range. The $S/N$ begins to plateau for $\ell\gtrsim7000$, and therefore our fiducial scale cut $\ell_{\rm max} = 10,000$ is sufficient to capture the sensitivity of future experiments to the different effects studied here.

  \subsubsection{Impact of satellite galaxies}  \label{sssec:results.satellites}

   \begin{figure*}
        \centering
        \includegraphics[width=0.485\linewidth]{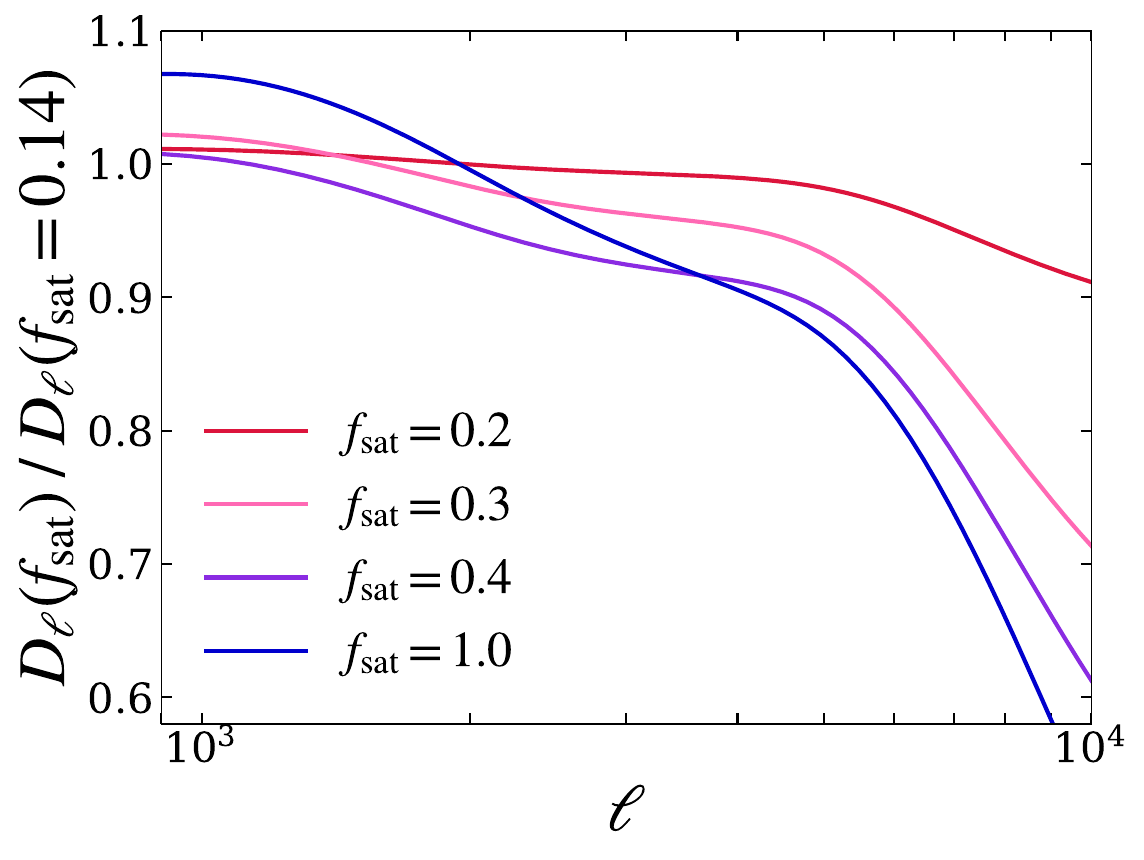}
        \includegraphics[width=0.49\linewidth]{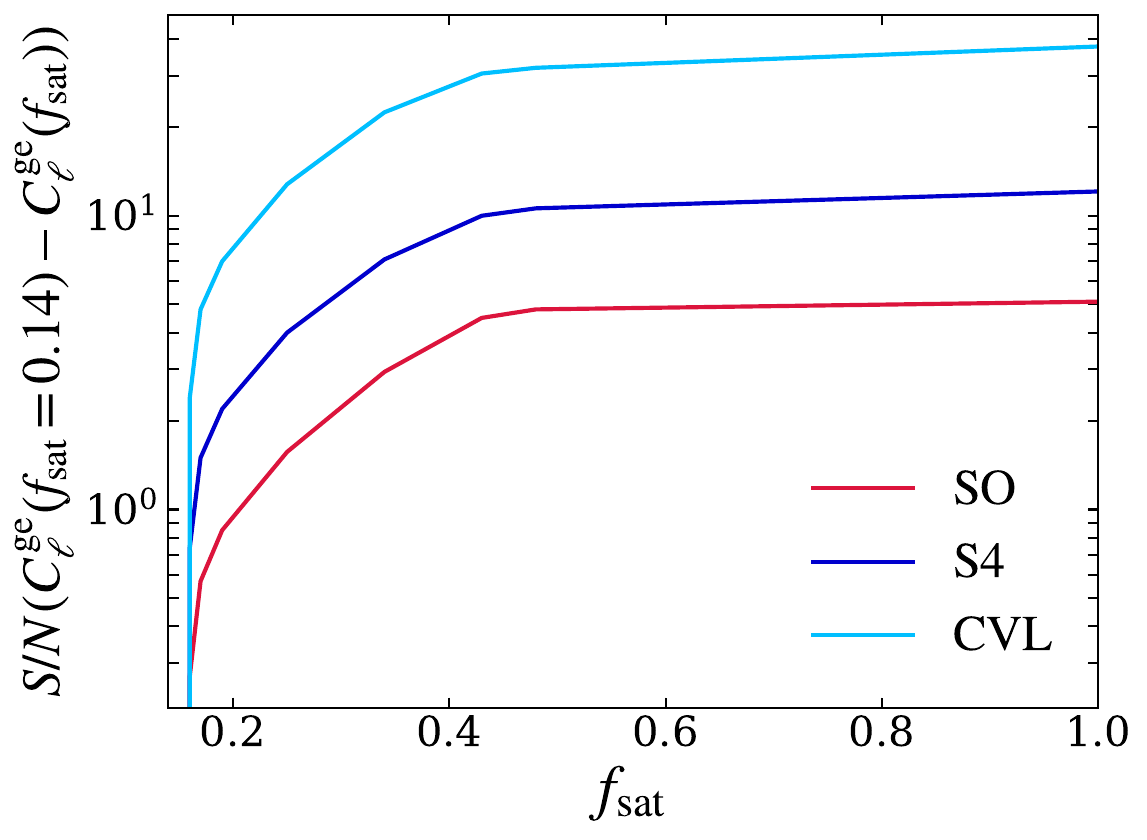}
        \caption{\textit{Left:} The ratio of the angular power spectrum of the perpendicular component between models that include satellite galaxies and the fiducial $(f_{\rm sat} = 0.14)$ case. The different lines correspond to HOD models with varying satellite fractions, as quoted in the legend. The large-scale galaxy bias is held fixed across all models to isolate the impact of satellites. The $\ell$-axis is restricted to small angular scales, where the satellite contribution becomes significant. \textit{Right:} The signal-to-noise ratio, $S/N$, as a function of the fraction of satellites, $f_{\rm sat}$, in the HOD model relative to the fiducial case. We consider three different noise levels: the Simons Observatory (red), CMB-S4 (dark blue), and the cosmic variance limit (light blue).}
       \label{fig:satellites}
    \end{figure*}

    In the left panel of Figure \ref{fig:satellites}, we illustrate how the inclusion of satellite galaxies in the electron-galaxy cross-power spectrum, $P_{\rm eg}(k)$, affects the angular power spectrum of the transverse $\rm \langle ge \rangle \langle mm \rangle$ component. For each galaxy sample, we fix the large-scale galaxy bias, $b_{\rm g}$, to isolate the impact of satellites. Under this constraint, the effect of satellites is limited primarily to small scales dominated by the one-halo term, allowing us to focus on their contribution to the $P_{\perp,1}$ term, which governs the total power spectrum on small scales. We find that incorporating satellite galaxies suppresses power on the smallest scales, driven by the fact that satellite galaxies trace the outer regions of the halo where the electron density is lower. The degree of suppression increases with the satellite fraction, $f_{\rm sat}$, of the galaxy sample. As shown in the right-hand panel of Figure \ref{fig:satellites}, this suppression propagates to the signal-to-noise ratio, $S/N$, which, for a given noise level, decreases as $f_{\rm sat}$ increases. The figure compares the $S/N$ of the fiducial sample -- based on the HOD parameters from \cite{Zhou:2020} with $f_{\rm sat} = 0.14$ -- to samples with higher satellite fractions. For a fixed noise level, the relative degradation in $S/N$ becomes more pronounced with increasing $f_{\rm sat}$, but plateaus for $f_{\rm sat} \gtrsim 0.45$. As in previous results, we observe an enhanced sensitivity to this effect for future experiments relative to SO.


   \subsubsection{Impact of gas model}\label{sssec:results.baryon_model}
    \begin{table}[h!]
      \centering
      \begin{tabular}{llcc}
        \hline
        Effect & Term & $S/N$ (HE) & $S/N$ (Battaglia) \\
        \hline
        \hline
        Total kSZ & $C_{\ell, \perp, \, \rm tot}^{\rm ge}$ & 34.2 & 96.4 \\
        Cross-term & $C_{\ell, \perp, \, 2}^{\rm ge}$ & 4.18 & 8.37 \\
        Satellite contribution & $C_{\ell, \perp,\, 1}^{\rm ge, \, \rm (cen+sats)}$ & 5.10 & 6.86 \\
        Two-halo contribution & $C_{\ell, \perp,\, 1}^{\rm ge, \, \rm (2h)}$ & 6.33 & 6.83 \\
        Connected trispectrum & $C_{\ell, \perp, \, \rm c}^{\rm ge}$ & 4.94 & 5.45  \\
        \hline
      \end{tabular}
      \caption{The signal-to-noise ratio, $S/N$, for the different contributions to the kSZ signal under two baryon models: the hydrostatic equilibrium (HE) profile and the Battaglia profile. Here, we use $\ell_{\rm max} = 10,000$ and the fiducial values of the baryonic parameters.}
      \label{tab:S/N_Battaglia}
    \end{table}
    The kSZ signal is directly sensitive to the effects of baryonic feedback, which governs the fraction of gas ejected by active galactic nuclei (AGN) outflows. Therefore, it is essential to assess how the choice of model for the gas distribution impacts the results presented here. Our analysis has thus far assumed an electron density profile based on the hydrostatic equilibrium (HE) model described in Section \ref{ssec:theory.implementation}, with a feedback strength (characterised primarily by the $\log_{10}M_{\rm c}$ parameter) compatible with current observations \cite{LaPosta:2025,2403.13794}. For comparison, we now consider the Battaglia profile from \cite{Battaglia:2016} as an alternative model.

\begin{figure*}
        \centering
        \includegraphics[width=0.5\linewidth]{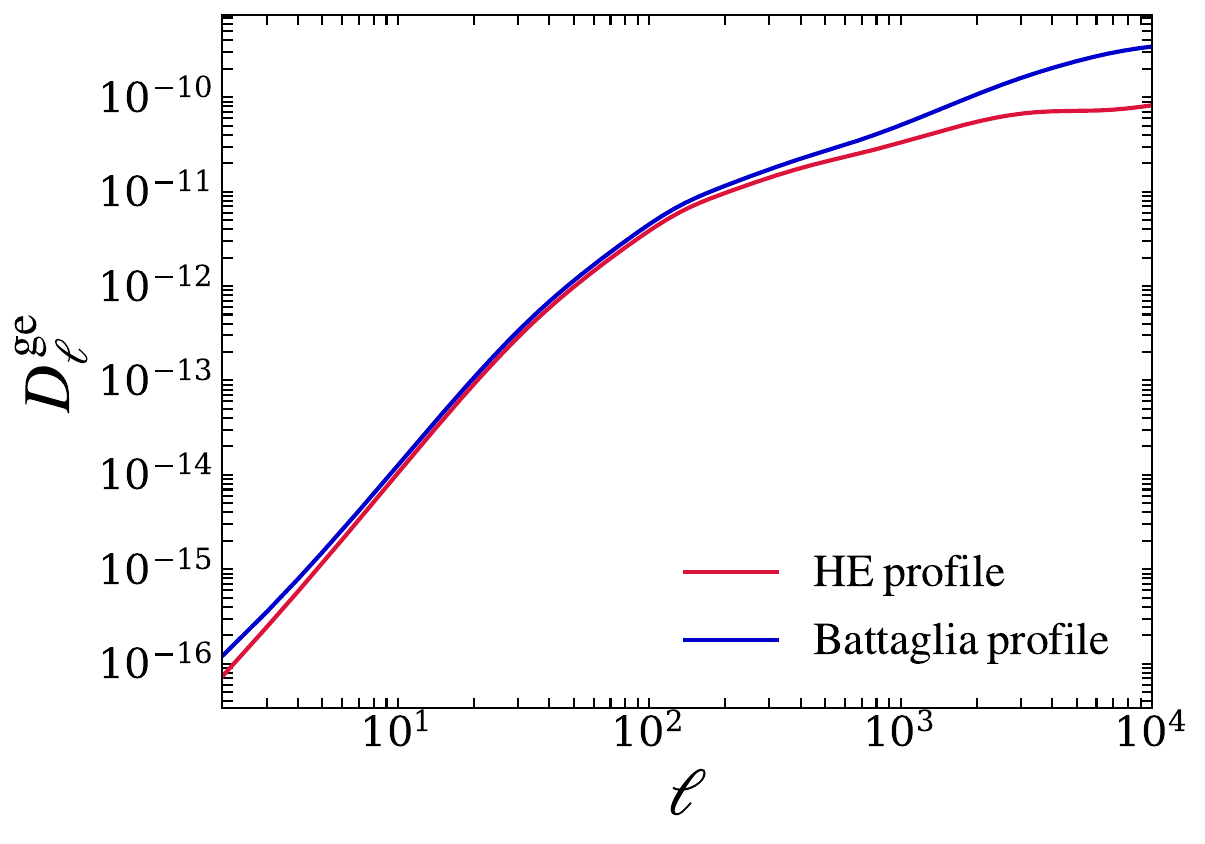}
        \includegraphics[width=0.49\linewidth]{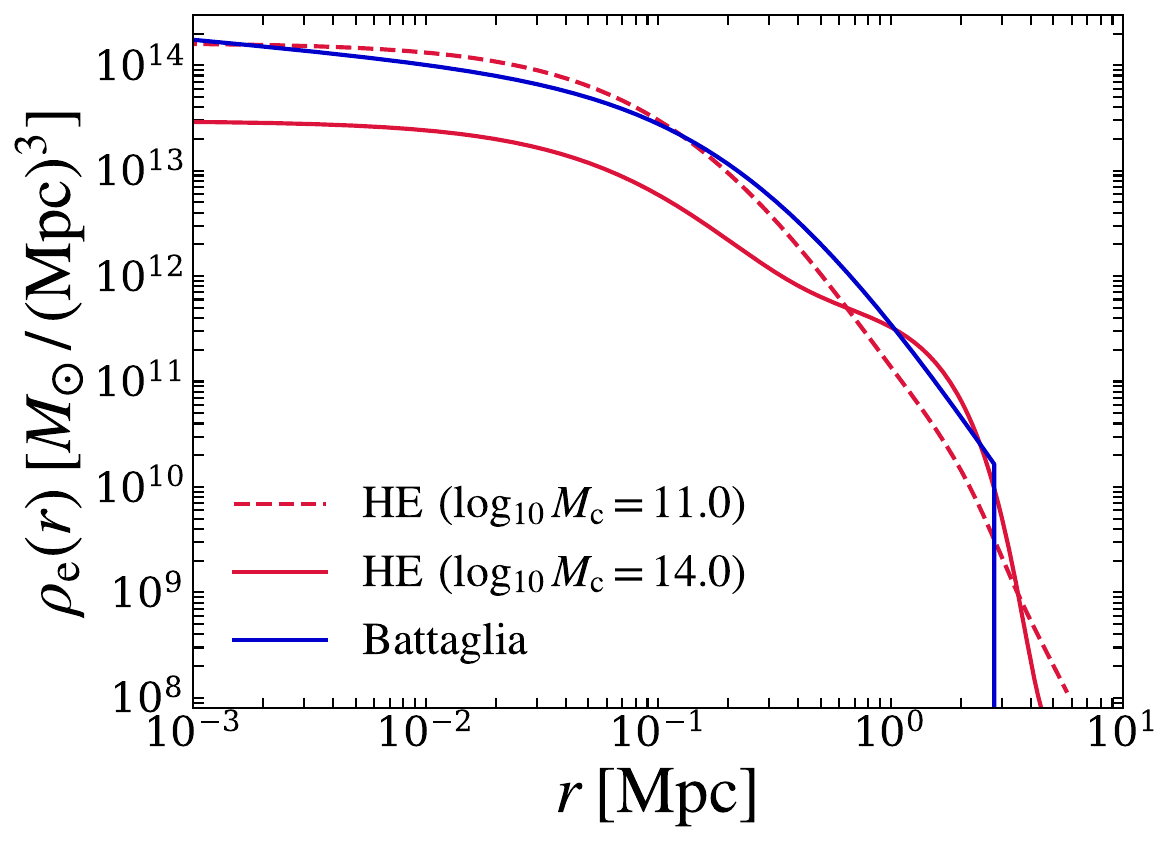}
        \caption{{\sl Left:} the impact of the baryon model on the angular power spectrum of the kSZ effect. The blue curve corresponds to the result assuming the hydrostatic equilibrium (HE) electron density profile, while the red curve shows the result using the Battaglia profile. {\sl Right:} the impact of baryonic feedback on the real-space electron density profiles as a function of comoving distance and for haloes of mass $M_{\rm halo} = 10^{13} M_{\odot}$. The HE profiles with low and high feedback strengths are shown as dashed and solid red lines, respectively, while the Battaglia profile is represented by the blue line.}\label{fig:baryon_model}
    \end{figure*}

    The left panel of Figure \ref{fig:baryon_model} shows the angular power spectrum obtained using the Battaglia profile, plotted relative to the result from the HE model. We observe that the Battaglia profile exhibits greater power on small angular scales. This enhancement can be attributed to two effects: its steeper slope at small scales and its weaker baryonic feedback relative to the HE model for the fiducial parameters used here. The impact of baryonic feedback is illustrated in the right panel of Figure \ref{fig:baryon_model}, which presents the real-space electron density profiles as a function of comoving distance from the halo centre for haloes of mass $M_{\rm halo} = 10^{13} M_{\odot}$ (roughly corresponding to the halos populated by the LRG sample considered here). Notably, we observe that the amplitude of the HE profile approaches that of the Battaglia profile on small scales when the baryonic feedback strength is reduced. Specifically, this is achieved by decreasing the feedback parameter $M_{\rm c}$ from its fiducial value of $\log_{10} M_{\rm c} = 14.0$ to $\log_{10} M_{\rm c} = 11.0$.

    The corresponding signal-to-noise ratios, computed assuming the baseline noise levels for SO, are presented in Table \ref{tab:S/N_Battaglia}. We find that using the Battaglia profile results in a significant increase in the $S/N$ of the total signal by a factor of $\sim3$. This difference highlights the sensitivity of the small-scale power spectrum to the choice of baryonic physics. Both the $S/N$ of the $\rm \langle gm\rangle\langle em\rangle$ term and the satellite galaxy contribution also increase significantly. This is reasonable, as both effects extend to the small scales modified by the Battaglia profile. In contrast, we find that the signal-to-noise ratios for the two-halo term and the connected trispectrum remain largely unchanged. This is consistent with expectations, as these contributions are primarily relevant on large scales, where both gas profiles yield the same total baryon mass when integrated out to large radii.

    Ultimately, the relative impact of the different secondary effects studied here depends on the specific gas distribution for the haloes probed by the galaxy sample under consideration. Current measurements tend to favour strong feedback scenarios \cite{2404.06098,Hadzhiyska:2024}, in which case, as discussed above, all the effects described here -- including the large-scale connected trispectrum and two-halo contributions -- will need to be included in the model for future experiments. In this context, it is worth noting that \cite{Park:2018baryonic} analysed the kSZ signal in the original \textit{Illustris} simulation \cite{Vogelsberger:2014} and showed that its overly strong AGN feedback implementation produces a pronounced suppression of the kSZ power spectrum at small scales.
   
   \subsubsection{Uncertainties from velocity reconstruction} \label{sssec:results.sensitivity.velrec}
    One limitation of our analysis lies in the assumption of perfect velocity reconstruction. In practice, the galaxy velocities are reconstructed from the measured galaxy overdensity, $\delta_{\rm g}$, often by inverting the linearised continuity equation \cite{1510.06442,1504.03677}. In this framework, assuming a linear bias and neglecting redshift-space distortions, the velocity field can be reconstructed from the galaxy overdensity via
    \begin{equation}
      {\bf v}({\bf k})=(aHf)\frac{i{\bf k}}{k^2}\frac{\tilde{\delta}_{\rm g}({\bf k})}{b_{\rm g}},
    \end{equation}
    where $b_{\rm g}$ is the linear galaxy bias, and $\tilde{\delta}_{\rm g}$ is a filtered version of the galaxy overdensity field (typically smoothed over a given scale). This procedure leads to two main effects:
    \begin{itemize}
      \item On small scales, the different clustering properties of galaxy and matter will lead to a mis-estimation in the shape of the velocity power spectrum.
      \item Non-linear effects, discarded in the linearised continuity equation, shot noise, and the loss of power due to smoothing, lead to a loss of correlation between the true and reconstructed velocities. This is exacerbated in the presence of large redshift errors (e.g. for photometric redshifts), which effectively smooth the galaxy density field further in the radial direction.
    \end{itemize}

\begin{figure*}
       \centering
       \includegraphics[width=0.49\linewidth]{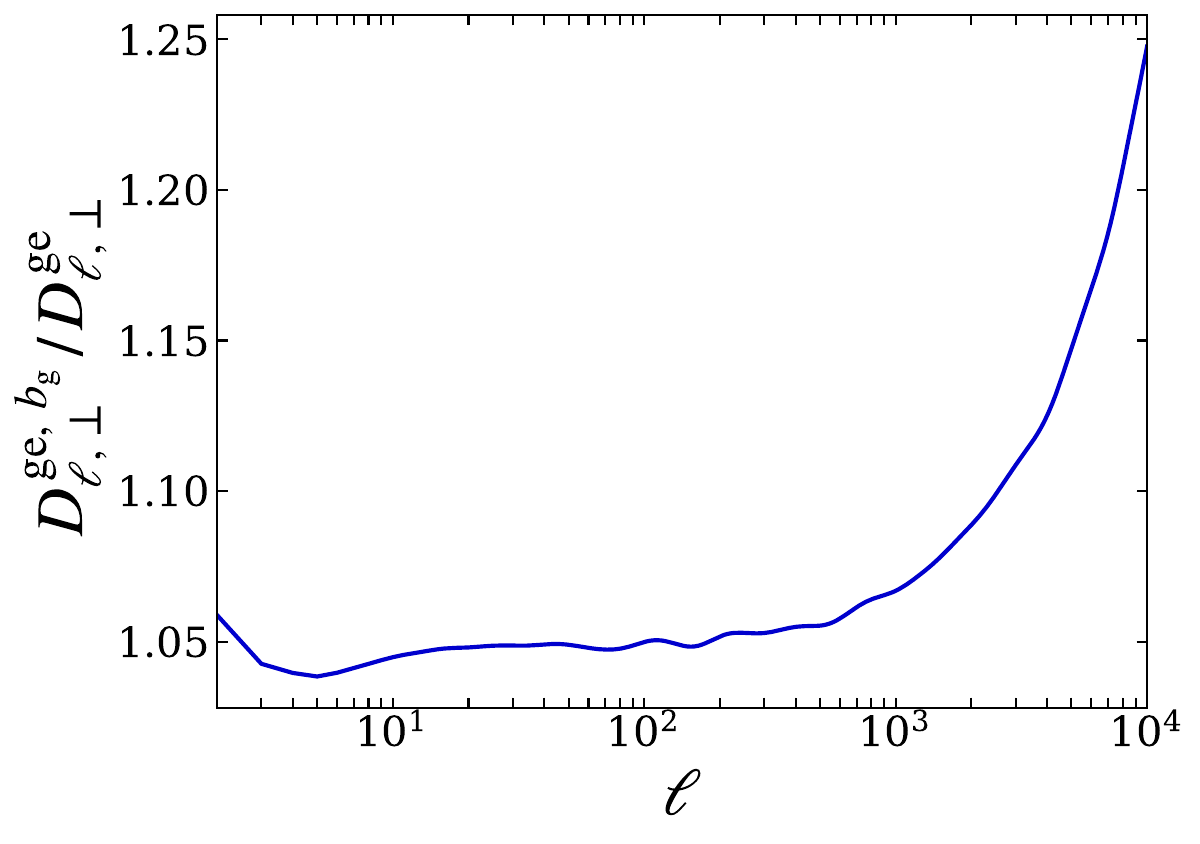}
       \includegraphics[width=0.46\linewidth]{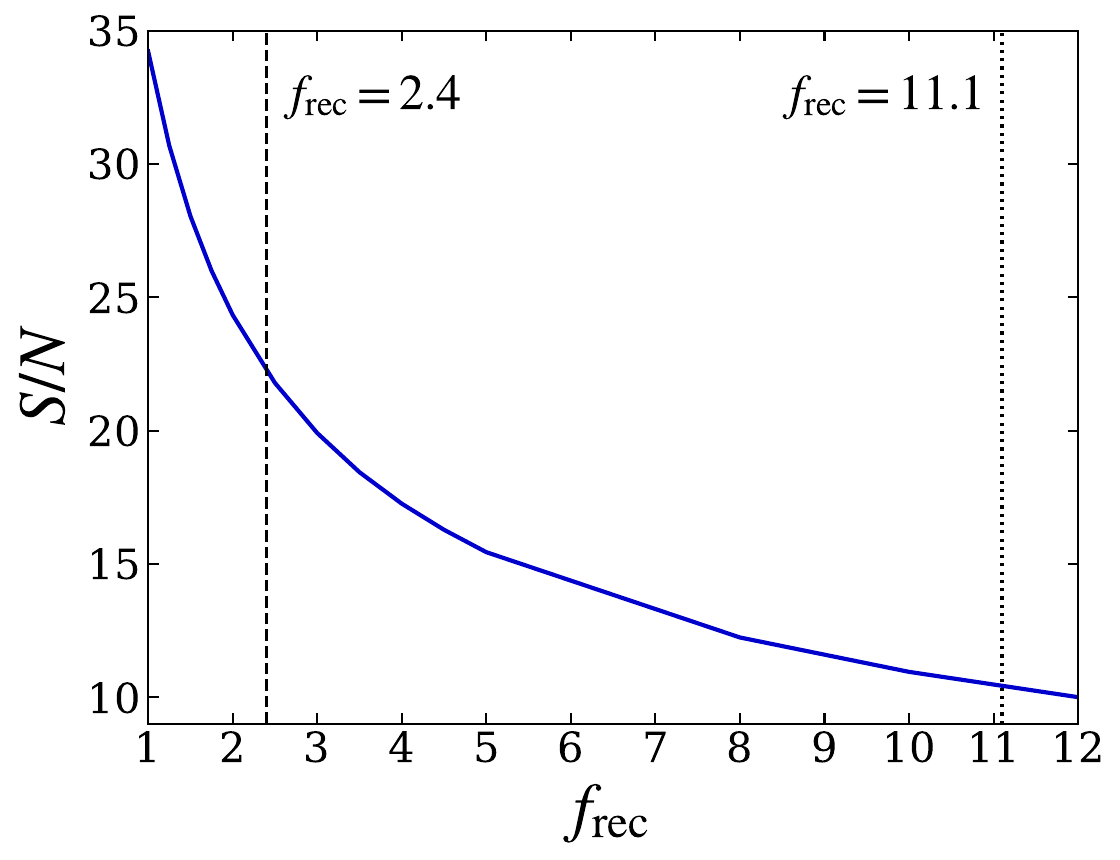}
       \caption{\textit{Left:} The impact of accounting for galaxy bias, $b_{\rm g}$, on the angular power spectrum for the transverse mode of the momentum field. We present the ratio of the angular power spectrum obtained when $b_{\rm g}$ is included in the matter–matter and matter–electron power spectra to that obtained when $b_{\rm g}$ is neglected. \textit{Right:} The signal-to-noise ratio, $S/N$, as a function of the velocity reconstruction parameter, $f_{\rm rec}$, which is related to the correlation coefficient through $f_{\rm rec} = 1/r^2$. The vertical lines denote the values of $f_{\rm rec}$ corresponding to different survey types. The dashed line at $f_{\rm rec} = 2.4$ corresponds to a spectroscopic survey, while the dotted line at $f_{\rm rec} = 11.1$ describes a photometric survey. We assume the baseline noise level of SO.}
       \label{fig:velocity_reconstruction}
    \end{figure*}

    The left panel in Figure \ref{fig:velocity_reconstruction} shows the impact of the first effect (incorrect small-scale dependence) on the kSZ power spectrum, estimated by simply replacing $P_{\rm mm}(k)$ in Equation \eqref{eq:P_perp_1} with $P_{\rm gm}/b_{\rm g}$. The effect is relatively small, modifying the overall $S/N$ of the signal by $\sim10\%$. This is not surprising, since the velocity power spectrum dominates on large scales, where the linear bias approximation is appropriate. 

    To quantify the impact of the loss of correlation due to reconstruction, we simply scale the effective shot noise level of the reconstructed momentum field ($N_\ell^{\rm gg}$ in Equation \eqref{eq:clgg}). In our $S/N$ calculation this is equivalent to scaling the amplitude of the cross-power spectrum by a correlation coefficient $r\equiv1/\sqrt{f_{\rm rec}}$ (see e.g. \cite{Schaan:2021}). The impact of velocity reconstruction on the total $S/N$, is illustrated in the right-hand panel of Figure \ref{fig:velocity_reconstruction}. The dashed line at $f_{\rm rec} = 2.4$ corresponds to the typical correlation achieved from spectroscopic surveys ($r = 0.64$), while the dotted line at $f_{\rm rec} = 11.1$ is more characteristic of a photometric LRG-like sample ($r = 0.3$) \cite{Hadzhiyska:2024}. Depending on the velocity reconstruction accuracy, the $S/N$ values reported here could decrease by up to a factor $\sim3$, although the relative contribution from the different effects listed in Table \ref{tab:S/N} would remain the same. Note also that the figures reported here correspond to the signal measured in a single galaxy sample and redshift bin. Current and future measurements are and will be carried out over a range of redshifts and galaxy types \cite{Hadzhiyska:2024,2507.14136,2503.19870,2412.03631}, boosting the significance of all signals explored here.

 \section{Conclusions} \label{sec:conclusions}
   Measurements of the stacked kSZ signals using velocity reconstruction from galaxies are an invaluable tool for cosmology and astrophysics, encoding key information to quantify the impact of feedback on the distribution of gas, and the associated suppression in the small-scale amplitude of the matter power spectrum. As CMB experiments improve in sensitivity, the need for accurate models of this signal becomes more pressing, and previously neglected effects may become significant.
   
   In this sense, current models of the stacked kSZ effect often overlook several contributions, including the two-halo term, the impact of satellite galaxies, and higher-order correlations between the velocity and density fields. In this work, we have presented a comprehensive analytical framework, deriving from first principles all terms, for both longitudinal and transverse modes, that contribute to the angular cross-correlation between the kSZ effect and the projected galaxy momentum field.

   Using forecasts for the sensitivity of upcoming CMB experiments, specifically the Simons Observatory and CMB-S4, we quantify the ability of such experiments to detect the various contributions to the kSZ-galaxy angular power spectrum. We find that the cross-term, describing the correlation of the galaxy and electron overdensities with the peculiar velocity field (Equation \eqref{eq:P_par_2}) will be detectable at the $\sim4$-$6\sigma$ level with these experiments. Interestingly, the connected non-Gaussian trispectrum term contributing to the transverse power spectrum could also be detectable, at the level of $\sim5$-$6\sigma$ and therefore cannot be neglected. Together, these terms characterise the correlation between the positions of galaxies (and of the gas) with the large-scale velocity field, which any theoretical model of the kSZ signal must therefore account for \cite{2412.09526}. Emulator-based approaches, for instance, will need to be built by reproducing the galaxy-kSZ cross-correlation directly, rather than simply measuring the electron density profile of haloes in the simulation. Reinforcing this conclusion, we also find that the two-halo term may become significant, at the $\sim6$-$7\sigma$ level. This contribution will be particularly relevant when tracing the density of the ejected gas over distances substantially larger than the halo virial radius. 

   We have also shown that the impact from satellite galaxies can be quite substantial, and could be detected at the $\sim5$-$10\sigma$ level. Since this effect leads to a suppression of power on the smallest scales, where sensitivity to the kSZ signal is most relevant, a correct model for the satellite contribution will be vital to obtain unbiased constraints on baryonic feedback models \cite{2505.20413}. We further quantified the impact of different models for the electron profile, used to describe the gas distribution in and around haloes. In particular, we compare the Battaglia profile to the hydrostatic equilibrium (HE) profile. We find that the signal-to-noise increases by a factor of $\sim 3$ when adopting the Battaglia profile relative to the HE profile. This emphasises the sensitivity of kSZ measurements to baryonic feedback. Finally, we find that the contribution from longitudinal modes is completely negligible, even when assuming an ideal cosmic variance-limited experiment, indicating that it is safe to neglect the parallel component altogether in kSZ stacking analyses.

   The analysis presented here is subject to a number of shortcomings, which could be addressed in future work. First, for simplicity we have assumed perfect velocity reconstruction, summarising its impact in terms of an overall velocity correlation coefficient $r$ (expressed in terms of $f_{\rm rec}$ in Section \ref{sssec:results.sensitivity.velrec}). The impact of imperfect reconstruction should be studied in greater detail, characterising the potential modifications not only to the overall amplitude of the measured signal, but also to its shape, which could be significant given future instrumental sensitivities. It will also be important to quantify the level to which the impact on the signal amplitude can be calibrated and corrected for, and whether uncertainties in this calibration can be reliably propagated. Our assessment of the relative impact of satellites has been relatively simplistic. Specifically, we have assumed that the presence of satellites is uncorrelated with the electron density distribution of the host halo, and we have ignored the impact of feedback on the subhalos populated by these satellites \cite{2412.09526}. More generally, the model used to describe the gas distribution in the presence of AGN feedback is rather crude, assuming a simple hydrostatic equilibrium profile for the bound gas, rather than more general parametrisations used in the literature \citep{Schneider:2015,Arico:2021}. This, as we have seen in Section \ref{sssec:results.baryon_model}, may have a strong impact on the $S/N$ estimates presented here. Furthermore, our reliance on the analytical halo model, means that our results are subject to some of the inaccuracies of this formalism \cite{Mead:2020,2009.01858}, including its under-estimation of the power spectrum on scales straddling the regimes dominated by the one-halo and two-halo terms.
   
   Overall, our results demonstrate that improved modelling of both velocity reconstruction and baryonic physics is essential for fully benefiting from the potential of next-generation CMB experiments in their sensitivity to the kSZ effect. In particular, our work reinforces the value of the kSZ effect as a sensitive tracer of both large-scale structure and baryonic feedback, offering new avenues for precision cosmology and astrophysics.

\acknowledgments
  We would like to thank Ra\'ul Angulo, Boris Bolliet, Boryana Hadzhiyska, and Emmanuel Schaan for useful comments and discussions. AW is supported by a Science and Technology Facilities Council studentship. DA and AP acknowledge support from the Beecroft Trust.

\bibliographystyle{JHEP}
\bibliography{references.bib}

\end{document}